\documentclass[final]{IEEEtran}
\usepackage{amsthm,amssymb,graphicx,multirow,amsmath,color,amsfonts,physics}%,ulem}
\usepackage[update,prepend]{epstopdf}
\usepackage[noadjust]{cite}
\usepackage{tikz}
\usepackage{bbm} % for \nbb1 
\usepackage{pdfpages}
\usepackage{balance}
\usepackage{multirow}
\usepackage{comment}
\usepackage{subfigure}
\usepackage{yhmath}
\newtheorem{corollary}{Corollary}
\allowdisplaybreaks % Allows breaking of eqnarray over multiple pages (avoids unnecessary blanks in the document before eqnarray)
\allowdisplaybreaks % Allows breaking of eqnarray over multiple pages (avoids unnecessary blanks in the document before eqnarray)
\usepackage{float}

\begin{document}
% Bold lowercase: syntax \nb# where # is {a ... z, 0,1}
\def\nba{{\mathbf{a}}}
\def\nbb{{\mathbf{b}}}
\def\nbc{{\mathbf{c}}}
\def\nbd{{\mathbf{d}}}
\def\nbe{{\mathbf{e}}}
\def\nbf{{\mathbf{f}}}
\def\nbg{{\mathbf{g}}}
\def\nbh{{\mathbf{h}}}
\def\nbi{{\mathbf{i}}}
\def\nbj{{\mathbf{j}}}
\def\nbk{{\mathbf{k}}}
\def\nbl{{\mathbf{l}}}
\def\nbm{{\mathbf{m}}}
\def\nbn{{\mathbf{n}}}
\def\nbo{{\mathbf{o}}}
\def\nbp{{\mathbf{p}}}
\def\nbq{{\mathbf{q}}}
\def\nbr{{\mathbf{r}}}
\def\nbs{{\mathbf{s}}}
\def\nbt{{\mathbf{t}}}
\def\nbu{{\mathbf{u}}}
\def\nbv{{\mathbf{v}}}
\def\nbw{{\mathbf{w}}}
\def\nbx{{\mathbf{x}}}
\def\nby{{\mathbf{y}}}
\def\nbz{{\mathbf{z}}}
\def\nb0{{\mathbf{0}}}
\def\nb1{{\mathbf{1}}}

% Bold capital letters: syntax \nb# where # is {A ... Z}
\def\nbA{{\mathbf{A}}}
\def\nbB{{\mathbf{B}}}
\def\nbC{{\mathbf{C}}}
\def\nbD{{\mathbf{D}}}
\def\nbE{{\mathbf{E}}}
\def\nbF{{\mathbf{F}}}
\def\nbG{{\mathbf{G}}}
\def\nbH{{\mathbf{H}}}
\def\nbI{{\mathbf{I}}}
\def\nbJ{{\mathbf{J}}}
\def\nbK{{\mathbf{K}}}
\def\nbL{{\mathbf{L}}}
\def\nbM{{\mathbf{M}}}
\def\nbN{{\mathbf{N}}}
\def\nbO{{\mathbf{O}}}
\def\nbP{{\mathbf{P}}}
\def\nbQ{{\mathbf{Q}}}
\def\nbR{{\mathbf{R}}}
\def\nbS{{\mathbf{S}}}
\def\nbT{{\mathbf{T}}}
\def\nbU{{\mathbf{U}}}
\def\nbV{{\mathbf{V}}}
\def\nbW{{\mathbf{W}}}
\def\nbX{{\mathbf{X}}}
\def\nbY{{\mathbf{Y}}}
\def\nbZ{{\mathbf{Z}}}

% \mathcal: syntax \ncal# where # is {A ... Z}
\def\ncalA{{\mathcal{A}}}
\def\ncalB{{\mathcal{B}}}
\def\ncalC{{\mathcal{C}}}
\def\ncalD{{\mathcal{D}}}
\def\ncalE{{\mathcal{E}}}
\def\ncalF{{\mathcal{F}}}
\def\ncalG{{\mathcal{G}}}
\def\ncalH{{\mathcal{H}}}
\def\ncalI{{\mathcal{I}}}
\def\ncalJ{{\mathcal{J}}}
\def\ncalK{{\mathcal{K}}}
\def\ncalL{{\mathcal{L}}}
\def\ncalM{{\mathcal{M}}}
\def\ncalN{{\mathcal{N}}}
\def\ncalO{{\mathcal{O}}}
\def\ncalP{{\mathcal{P}}}
\def\ncalQ{{\mathcal{Q}}}
\def\ncalR{{\mathcal{R}}}
\def\ncalS{{\mathcal{S}}}
\def\ncalT{{\mathcal{T}}}
\def\ncalU{{\mathcal{U}}}
\def\ncalV{{\mathcal{V}}}
\def\ncalW{{\mathcal{W}}}
\def\ncalX{{\mathcal{X}}}
\def\ncalY{{\mathcal{Y}}}
\def\ncalZ{{\mathcal{Z}}}

% \mathbb: syntax \nbb# where # is {A ... Z}
\def\nbbA{{\mathbb{A}}}
\def\nbbB{{\mathbb{B}}}
\def\nbbC{{\mathbb{C}}}
\def\nbbD{{\mathbb{D}}}
\def\nbbE{{\mathbb{E}}}
\def\nbbF{{\mathbb{F}}}
\def\nbbG{{\mathbb{G}}}
\def\nbbH{{\mathbb{H}}}
\def\nbbI{{\mathbb{I}}}
\def\nbbJ{{\mathbb{J}}}
\def\nbbK{{\mathbb{K}}}
\def\nbbL{{\mathbb{L}}}
\def\nbbM{{\mathbb{M}}}
\def\nbbN{{\mathbb{N}}}
\def\nbbO{{\mathbb{O}}}
\def\nbbP{{\mathbb{P}}}
\def\nbbQ{{\mathbb{Q}}}
\def\nbbR{{\mathbb{R}}}
\def\nbbS{{\mathbb{S}}}
\def\nbbT{{\mathbb{T}}}
\def\nbbU{{\mathbb{U}}}
\def\nbbV{{\mathbb{V}}}
\def\nbbW{{\mathbb{W}}}
\def\nbbX{{\mathbb{X}}}
\def\nbbY{{\mathbb{Y}}}
\def\nbbZ{{\mathbb{Z}}}

% \mathfrak:
\def\nfrakR{{\mathfrak{R}}}

% Roman: {\rm } syntax \nrm# where # is {a ... z}
\def\nrma{{\rm a}}
\def\nrmb{{\rm b}}
\def\nrmc{{\rm c}}
\def\nrmd{{\rm d}}
\def\nrme{{\rm e}}
\def\nrmf{{\rm f}}
\def\nrmg{{\rm g}}
\def\nrmh{{\rm h}}
\def\nrmi{{\rm i}}
\def\nrmj{{\rm j}}
\def\nrmk{{\rm k}}
\def\nrml{{\rm l}}
\def\nrmm{{\rm m}}
\def\nrmn{{\rm n}}
\def\nrmo{{\rm o}}
\def\nrmp{{\rm p}}
\def\nrmq{{\rm q}}
\def\nrmr{{\rm r}}
\def\nrms{{\rm s}}
\def\nrmt{{\rm t}}
\def\nrmu{{\rm u}}
\def\nrmv{{\rm v}}
\def\nrmw{{\rm w}}
\def\nrmx{{\rm x}}
\def\nrmy{{\rm y}}
\def\nrmz{{\rm z}}

% Special symbols
\def\nbydef{:=}
\def\nborel{\ncalB(\nbbR)}
\def\nboreld{\ncalB(\nbbR^d)}
\def\sinc{{\rm sinc}}

% Theorems etc.
\newtheorem{lemma}{Lemma}
\newtheorem{thm}{Theorem}
\newtheorem{definition}{Definition}
\newtheorem{ndef}{Definition}
\newtheorem{nrem}{Remark}
\newtheorem{theorem}{Theorem}
\newtheorem{prop}{Proposition}
\newtheorem{cor}{Corollary}
\newtheorem{example}{Example}
\newtheorem{remark}{Remark}
\newtheorem{assumption}{Assumption}
	
%%%%%%%% Backwards compatibility

\newcommand{\ceil}[1]{\lceil #1\rceil}
\def\argmin{\operatorname{arg~min}}
\def\argmax{\operatorname{arg~max}}
\def\figref#1{Fig.\,\ref{#1}}%
\def\E{\mathbb{E}}
\def\EE{\mathbb{E}^{!o}}
\def\P{\mathbb{P}}
\def\pc{\mathtt{P_c}}
\def\rc{\mathtt{R_c}}   % rate coverage
\def\p{p}

\def\V{\operatorname{Var}}
\def\erfc{\operatorname{erfc}}
\def\erf{\operatorname{erf}}
\def\opt{\mathrm{opt}}
\def\R{\mathbb{R}}
\def\Z{\mathbb{Z}}

\def\LL{\mathcal{L}^{!o}}
\def\var{\operatorname{var}}
\def\supp{\operatorname{supp}}

\def\N{\sigma^2}
\def\T{\beta}							% Threshold = \beta_i
\def\sinr{\mathtt{SINR}}			% Signal to interference plus noise ratio
\def\snr{\mathtt{SNR}}
\def\sir{\mathtt{SIR}}
\def\ase{\mathtt{ASE}}
\def\se{\mathtt{SE}}

\def\calN{\mathcal{N}}
\def\FE{\mathcal{F}}
\def\calA{\mathcal{A}}
\def\calK{\mathcal{K}}
\def\calT{\mathcal{T}}
\def\calB{\mathcal{B}}
\def\calE{\mathcal{E}}
\def\calP{\mathcal{P}}
\def\calL{\mathcal{L}}
%\DeclareMathOperator{\Tr}{Tr}
%\DeclareMathOperator{\rank}{rank}
%\DeclareMathOperator{\Pois}{Pois}

%\DeclareMathOperator{\TC}{\mathtt{TC}}
%\DeclareMathOperator{\TCL}{\mathtt{TC_l}}
%\DeclareMathOperator{\TCU}{\mathtt{TC_u}}

% Fading
\def\l{\ell}
\newcommand{\fad}[2]{\ensuremath{\mathtt{h}_{#1}[#2]}}
\newcommand{\h}[1]{\ensuremath{\mathtt{h}_{#1}}}

\newcommand{\err}[1]{\ensuremath{\operatorname{Err}(\eta,#1)}}
\newcommand{\FD}[1]{\ensuremath{|\mathcal{F}_{#1}|}}

%% Symbols changed
% \def\i{\mathbf{1}}					% changed to \nb1
% \def\d{\mathrm{d}}					% changed to \nrmd
% \def\L{\mathcal{L}}					% changed to \ncalL
% \begin{definition}					% changed to \begin{ndef}

% \l also gives problems. Use \ell after defining it if needed.

%% D2D def
\def\Bx{{\mathcal{B}}^x}
\def\Bxx{{\mathcal{B}}^{x_0}}
\def\jx{y}
\def\m{(\bar{n}-1)}
\def\mm{\bar{n}-1}
\def\Nx{{\mathcal{N}}^x}
\def\Nxo{{\mathcal{N}}^{x_0}}
\def\wj{w_{jx_0}}
\def\uij{u_{jx}}
 \def\yj{y}
 \def\yjx{y}
 \def\zjx{z_x}
 \def \tx {y_0}
 \def \htx {h_0}

\def\rx{z_{1}}
\def\ry{z_{2}}

\def\Rx{Z_{1}}
\def\Ry{Z_{2}}

%% fading
\def \hyxx {h_{y_{x_0}}}
\def \hyx {h_{y_x}}

\def\nbb1{\mathbbm{1}}
\def\xi{\textbf{x}_i}
\def\xj{\textbf{x}_j}
\def\xx{\textbf{x}_0}
\def\yi{\textbf{y}_i}
\def\yj{\textbf{y}_j}
\def\yy{\textbf{y}_0}
\def\oe{\textbf{o}_e}
\def\wik{\textbf{w}_{i,k}}
\def\ie{{\em i.e. }}
\def\eg{{\em e.g. }}
\def\iid{{\em i.i.d. }}
\def\avg{\rm avg}

\def\rmnuma{\rm\uppercase\expandafter{\romannumeral1}}
\def\rmnumb{\rm\uppercase\expandafter{\romannumeral2}}
\def\rmnumc{\rm\uppercase\expandafter{\romannumeral3}}
\def\rmnumd{\rm\uppercase\expandafter{\romannumeral4}}
\def\rmnume{\rm\uppercase\expandafter{\romannumeral5}}
\def\rmnumf{\rm\uppercase\expandafter{\romannumeral6}}
\pagenumbering{gobble}
\graphicspath{{./Figures/}}
\title{
Connectivity of LEO Satellite Mega Constellations: An Application of Percolation Theory on a Sphere}
\author{
 Hao Lin,  Mustafa A. Kishk and Mohamed-Slim Alouini
\thanks{Hao Lin is with the Electrical and Computer Engineering Program, CEMSE Division, King Abdullah University of Science and Technology (KAUST),
Thuwal 23955-6900, Saudi Arabia (e-mail: hao.lin.std@gmail.com).\\
\indent Mustafa A. Kishk is with the Department of Electronic Engineering,
Maynooth University, Maynooth, W23 F2H6 Ireland (e-mail:
mustafa.kishk@mu.ie).\\
\indent Mohamed-Slim Alouini is with the CEMSE Division, King Abdullah
University of Science and Technology (KAUST), Thuwal 23955-6900,
Saudi Arabia (e-mail: slim.alouini@kaust.edu.sa).}
}

\maketitle
\vspace{-2cm}
\begin{abstract}
With the advent of the 6G era, global connectivity has become a common goal in the evolution of communications, aiming to bring Internet services to more unconnected regions. Additionally, the rise of applications such as the Internet of Everything and remote education also requires global connectivity. Non-terrestrial networks (NTN), particularly low earth orbit (LEO) satellites, play a crucial role in this future vision. Although some literature already analyze the coverage performance using stochastic geometry, the ability of generating large-scale continuous service area is still expected to analyze.  Therefore, in this paper, we mainly investigate the necessary conditions of LEO satellite deployment for large-scale continuous service coverage on the earth. Firstly, we apply percolation theory to a closed spherical surface and define the percolation on a sphere for the first time. We introduce the sub-critical and super-critical cases to prove the existence of the phase transition of percolation probability. Then, through stereographic projection, we introduce the tight bounds and closed-form expression of the critical number of LEO satellites on the same constellation. In addition, we also investigate how the altitude and maximum slant range of LEO satellites affect percolation probability, and derive the critical values of them. Based on our findings, we provide useful recommendations for companies planning to deploy LEO satellite networks to enhance connectivity.
\end{abstract}
% \vspace{-0.3cm}
\begin{IEEEkeywords}
% \vspace{-0.3cm}
Percolation theory, LEO satellites, non-terrestrial networks, stereographic projection.
\end{IEEEkeywords}

\section{Introduction} \label{sec:Intro}

As a key technology of next-generation communications, non-terrestrial networks (NTN) have been proposed to enhance high-capacity global connectivity \cite{giordani2020non}. 3GPP Release 17 defined the New Radio (NR) to support NTN broadband Internet services, especially for rural and remote areas \cite{3GPPRelease17}. Narrowband Internet of Things (NB-IoT) over NTN has been preliminarily standardised and commercial deployments are ongoing, where low earth orbit (LEO) satellite constellations can be a solution to provide low latency service in a low cost \cite{3GPPRelease18,liberg2020narrowband}. Furthermore, in 3GPP Release 18 and 19, NTNs including LEO satellites aim to support regenerative payloads, and coverage and mobility enhancements, for more requirements from handheld terminals, NB-IoT and enhanced machine-type communication (eMTC) \cite{10500741}. They can help not only support 5G NR but also pave the way towards 6G technologies. LEO-satellite access networks have been deployed to provide seamless massive access and connect the unconnected areas on the earth \cite{xiao2022leo}. Examples of currently deployed or planned LEO satellite constellations include Starlink, OneWeb and Kuiper \cite{osoro2021techno,voicu2024handover}. Therefore, it is necessary to capture the ability of LEO satellites to enhance global connectivity.  \\
\indent Stochastic geometry is an important tool to evaluate the performance of large-scale wireless networks without losing accuracy \cite{haenggi2012stochastic}. It is widely used to evaluate the coverage performance of 2D or 3D wireless networks. For global coverage, it can also provide a basic geometric framework, especially using the binomial point process (BPP) and Poisson point process (PPP). Global connectivity is a vital performance indicator of large-scale wireless networks, where graph theory and percolation theory can help capture such performance metric \cite{haenggi2009stochastic,haenggi2012stochastic}. Percolation probability represents the probability of generating large-scale connected components in a network, where the nodes can be connected through multi-hops. Such an indicator can be used to capture a network's connectivity, robustness, cyber-security and path exposure \cite{elsawy2023tutorial}.\\
\indent Therefore, it is necessary to evaluate the connectivity of LEO satellites' coverage, that is, the ability to generate large-scale continuous paths on the earth that are covered by LEO satellites, through percolation theory. However, conventional percolation on wireless networks relies on a 2D plane or a 3D system, which is different from the realistic deployment of user equipment (UEs) or IoT devices on the earth and the coverage regions of LEO satellites. So that, the percolation analysis on the 3D sphere is still an unsolved problem.

\subsection{Related Work}
 In this paper, we apply percolation theory to a sphere for the first time to discuss the ability to generate large-scale continuous coverage paths of LEO satellites. Therefore, we divide the related works into: i) LEO satellite communications based on stochastic geometry and ii) percolation theory applications on wireless networks.\\%i) LEO satellite constellations for next-generation mobile communications and ii) percolation theory applications on wireless networks.\\
\indent \textit{LEO satellite communications based on stochastic geometry}: In recent years, LEO satellite communication has been a focus of next-generation communication technology. Stochastic geometry is widely used to evaluate the performance of communication systems with LEO satellites. In \cite{talgat2020stochastic}, authors studied the coverage performance of LEO satellite communication system, where satellite gateways on the ground act as relays between users and satellites. Authors in \cite{al2021analytic} and \cite{al2021optimal} evaluated the average downlink success probability for dense satellite networks and optimal satellite constellation altitude. Authors in \cite{al2022optimal} extended the work and investigated the optimal beamwidth and altitude for maximal uplink coverage of satellite networks. Authors in \cite{okati2020downlink} and \cite{okati2022nonhomogeneous} evaluated the average data rate and coverage probability using BPP and nonhomogeneous PPP, respectively. In \cite{okati2023stochastic}, they derived and verified the coverage probability of a multi-altitude LEO network. In \cite{park2022tractable}, authors derived the tight lower bound of coverage probability and found out the relationship between optimal average number of satellites and the altitude of satellites. A tractable framework was developed in \cite{salem2023exploiting} to evaluate the performance of downlink hybrid terrestrial/satellite networks in rural areas. Authors in \cite{shang2023coverage} derived the joint distance distribution of cooperative LEO satellites to the typical user, and obtained the optimal satellite density and satellite altitude to maximize the coverage probability. For Space-air-ground integrated networks (SAGIN), authors in \cite{yuan2023joint} proposed a simulated annealing algorithm-based optimization algorithm to optimize THz and RF channel allocation. Authors in \cite{wang2022ultra} studied the influence of gateway density and the setting of satellites constellations on latency and coverage probability. They established an optimization algorithm to maximize reliability and minimize latency, and obtained the ideal upper bound for these performances in \cite{wang2024ultra}. For a heterogeneous satellite network, authors in \cite{choi2024modeling} proved that the open access scenario can obtain a higher coverage probability than the closed access. Authors in \cite{choi2024analysis} investigated the key performance indices of a delay-tolerant data harvesting architecture, including the CDF of delay and harvest capacity. For different communication scenarios, authors in \cite{talgat2024stochastic} analyzed the uplink performance of IoT over LEO satellite communication with reliable coverage. An adaptive coverage enhancement (ACE) method was proposed in \cite{hong2024narrowband} for random access parameter configurations under diverse applications. Authors in \cite{bliss2024orchestrating} investigated the impact of onboard energy limitation, minimum elevation angle on downlink steady-state probability and availability. Authors in \cite{taojoint} proposed a throughput optimization algorithm for LEO satellite-based IoT networks and derived the closed-form expression of the throughput when LEO satellites are equipped with capture effect (CE) receiver and successive interference cancellation (SIC) receiver, respectively.

\indent \textit{Percolation theory applications on wireless networks}: Percolation theory and graph theory are widely used to evaluate the connectivity of large-scale networks, including multi-hops links, detective paths, continuous coverage, security, to name a few \cite{haenggi2009stochastic,haenggi2012stochastic,elsawy2023tutorial}. Authors in \cite{anjum2019percolation} modeled the homogeneous wireless balloon network (WBN) as a Gilbert disk model (GDM) and modeled the heterogeneous WBN as a Random Gilbert disk model (RGDM).
They derived the bounds of the critical node density of such WBNs. In \cite{anjum2020coverage}, they also derived the critical density of unmanned aerial vehicles (UAVs) to ensure the network coverage of UAV networks (UN). Using percolation theory, authors in \cite{wang2019cooperative} derived the critical density of camera sensors in clustered 3D wireless camera sensor networks (WCSN). Authors in \cite{zhaikhan2020safeguarding} characterized the critical density of spatial firewalls to prevent malware epidemics in large-scale wireless networks (LSWN). In \cite{yemini2019simultaneous}, authors established a model for the coexistence of random primary and secondary cognitive networks and proved the feasibility of the simultaneous connectivity. Based on dynamic bound percolation, authors in \cite{han2024dynamic} characterized the reliable topology evolution and proved that the dynamic topology evolution (DTE) model can improve the overall network performance.
In \cite{wu2023connectivity}, authors investigated the connectivity of large-scale reconfigurable intelligent surface (RIS) assisted integrated access and backhaul (IAB) networks.
Considering the directional antenna, authors in \cite{zhu2023connectivity} analyzed the connectivity of networks assisted by transmissive RIS.

\subsection{Contributions}
The contributions of this paper can be summarized as follows:\\
\indent \textit{A new perspective of connectivity of LEO satellite coverage.} In this paper, we evaluate the connectivity of LEO satellites' coverage using percolation theory. We adopt the percolation probability to show the ability to form giant continuous LEO satellite service areas or footprints for users on the earth. Such performance metric describes whether the devices that are moving on the earth can achieve continuous service and whether the Internet of Everything on a large scale can be supported by LEO satellites. \\
\indent \textit{Percolation theory on the Sphere.} In this paper, we first define the face percolation on the sphere using its stereographic projection onto a plane. We introduce the sub-critical and super-critical cases to investigate the lower and upper bounds of critical number of LEO satellites for percolation. Through stereographic projection, we derive the tight bounds for the critical threshold of the number of LEO satellites. By discussing the relationship between continuous percolation and discrete percolation, we obtain the closed-form expression of the critical number of LEO satellites. In addition, we also derive the closed-form critical expressions of altitude and maximum slant range of LEO satellites.\\
% \indent \textit{Suggestions for LEO satellite operators.} Through the percolation probability and coverage probability, we give different suggestions for different LEO satellite operators of different sizes. 

\section{System Model} \label{sec:SysMod}
% \begin{figure*}[htbp]
%     \centering
% \includegraphics[width=0.75\linewidth]{Figures/}
%     \caption{...}
%     \label{fig:...}
% \end{figure*}

For ease of tractability, the earth's surface is commonly considered a standard sphere with a radius $r_e=6371\, \rm km$, where the center of the earth is defined as $\oe$. The North Pole is located at $\textbf{w}_N$ and the South Pole is located at $\textbf{w}_S$. We assume that LEO satellites are uniformly distributed on a sphere at an altitude of $h$ from the ground, that is, the sphere centered at $\oe$ with a radius $r_s=r_e+h$. The locations of LEO satellites follow a BPP $\Phi=\{\yi\}$ with the number of satellites $N$, where $\yi$ represents the location of any satellite  \cite{talgat2020stochastic,wang2023coverage,talgat2024stochastic}. Notice that current LEO satellite constellations adopt different orbital design schemes, where Antarctic and Arctic region have different satellite densities from other regions. However, such a BPP assumption describes the future LEO satellite mega constellations with a massive number of satellites around the earth. Through antenna array and beamforming technique, each satellite can serve the users within the transmission angle $\eta$, \ie the nadir angle. The LEO satellite located at $\yi$ can provide the communication service for its spherical coverage area $\mathcal{A}_i=\mathcal{A}(\yi,\eta)$, which is defined as:
\begin{equation}
    \mathcal{A}(\yi,\eta)=\{\wik:\|\wik-\oe\|=r_e,\,\angle \oe\yi\wik\leq\eta\}.
\end{equation}
We can also define the center of the coverage area as $\xi$ so that such a coverage area can be defined using another symbol $\mathcal{O}_i=\mathcal{O}(\xi,\gamma)$, where:
\begin{equation}
    \mathcal{O}(\xi,\gamma)=\{\wik:\|\wik-\oe\|=r_e,\,\angle \xi\oe\wik\leq\gamma\},
\end{equation}
where $\gamma$ is the coverage angle of LEO satellites on the earth. Notice that $\mathcal{O}_i$ and $\mathcal{A}_i$ both represent the coverage area of the same satellite, they must satisfy
\begin{equation}
    \mathcal{O}(\xi,\gamma)=\mathcal{A}(\yi,\eta).
\end{equation}
Considering a user located at the boundary of LEO satellite's coverage, the distance to the satellite is called the maximum slant range $d_m$. The geometric relationships between $\gamma$, $\eta$, $h$ and $d_m$ are shown in the Fig. \ref{fig:etagamma} and Lemma \ref{lem:etagamma} in the following section.\\
\indent In this paper, we aim to analyze the connectivity of coverage areas of LEO satellites. Percolation theory, has its unique advantage of analyzing the connectivity, especially in a 2D plane. However, the definition of percolation on a sphere is less common in literature. Therefore, based on graph theory, we first define the connectivity of satellites' coverage areas as a 3D random graph $G_x(V_x,E_x)$, where $V_x=\{\xi\}$ is the set of locations of coverage centers and $E_x$ is edge set that shows whether the coverage areas of the considered satellites are connected. The edge set $E_x$ can be expressed as:
\begin{equation}
    E_x=\left\{\overline{\xi\xj}:    
    \left\{\begin{matrix}
 \angle \xi\oe\xj\leq 2\gamma \,\\
{\rm or}\\
\wideparen{\xi\xj}\leq 2r_e \gamma \, \\
{\rm or}\\
\|\xi\xj\|\leq 2r_e\sin\gamma

\end{matrix}\right\}\right\},
\end{equation}
where $\wideparen{\xi\xj}$ and $\|\xi\xj\|$ represent the spherical distance and Euclidean distance, respectively, between $\xi$ and $\xj$.\\
\indent In this paper, we propose to project the earth's surface onto a 2D plane which is tangent to the earth on the South Pole $\textbf{w}_S$. The random graph on the projection plane, which corresponds to $G_x(V_x,E_x)$, is defined as $G_z(V_z,E_z)$, where $V_z$ is the vertex set and $E_z$ is the edge set. We let $K_z\subset G_z(V_z,E_z)$ denote a connected component inside $G_z$ and let $K_z(0)$ denote the connected component covering the origin $\textbf{o}_z$ on the projection plane. On a 2D plane, percolation probability is commonly defined as the probability of generating a giant component whose set cardinality is infinite, \ie $\mathbb{P}\{|K_z|=\infty\}$. In our system, we define the percolation probability of LEO satellite coverage areas as a function of the number of satellites $N$ and the coverage angle $\gamma$, that is
\begin{equation}
    \theta(N,\gamma)=\mathbb{P}\{|K_z(0)|=\infty\},
\label{defineper}
\end{equation}
where the 2D connected component $K_z$ on a plane is projected from the 3D connected component $K_x$. Therefore, considering a fixed coverage angle $\gamma$ of each LEO satellite, the design objective of the considered system is:
\begin{equation}
    \begin{array}{ll}
       \text{ minimize}  & N  \\
       \text{ subject to}  & \theta(N,\gamma)>0.  \\
    \end{array}
    \label{design1}
\end{equation}
Similarly, considering a fixed number of LEO satellites deployed on the same altitude, we can also formulate the design objective as:
\begin{equation}
    \begin{array}{ll}
       \text{ minimize}  & \gamma  \\
       \text{ subject to}  & \theta(N,\gamma)>0. \\
    \end{array}
    \label{design2}
\end{equation}
It is worth noting that, the coverage angle $\gamma$ depends on the altitude $h$, maximum slant range $d_m$ or nadir angle $\eta$.\\
\indent In conclusion, using the tools of percolation theory, we mainly study the necessary conditions to form large-scale connected coverage areas on the earth. For ease of reading, we summarize most of the symbols in Table \ref{tab:TableOfNotations}.

\begin{table*}[htbp]
\caption{Table of Notations}
\centering
\begin{center}
\resizebox{\textwidth}{!}{
\renewcommand{\arraystretch}{1}%1.4
    \begin{tabular}{ {c} | {l} }
    \hline
        \hline
    \textbf{Notation} & \textbf{Description} \\ \hline
    $\textbf{o}_e$; $r_e$; $\textbf{w}_N$; $\textbf{w}_S$ & The center of earth; the radius of earth; the North Pole of earth; the South Pole of earth \\ \hline
    $h$; $r_s$; $N$ & The altitude of LEO satellites; the radius of LEO satellites' orbit; the number of LEO satellites\\ \hline
    $\Phi$; $\textbf{y}_i$; $\textbf{x}_i$ & The set of LEO satellites' locations; the location of the $i$th LEO satellite; the projection of the $i$th LEO satellites on the earth\\ \hline
    $\eta$; $\gamma$; $\epsilon$ & The nadir angle of LEO satellites; the coverage angle on the earth; the minimum elevation angle of users\\ \hline
    $p_{\rm cov}$; $p_{\rm ncov}$& The probability of each point on the earth being covered; the probability of each point on the earth being not covered \\ \hline
     $V_x$; $E_x$& The set of vertices (\ie coverage centers); the edge set which represents the connection between coverage areas\\ \hline
    $G_x(V_x,E_x)=\{V_x,E_x\}$& The random graph containing the vertex set $V_x$ and edge set $E_x$ on the earth (sphere)\\ \hline
    $G_z(V_z,E_z)=\{V_z,E_z\}$& The corresponding random graph of $G_x$ on the projection plane\\ \hline
    $K_x$; $K_x(0)$& The connected component on the sphere; the connected component on the sphere containing the South Pole $w_S$\\ \hline
    $K_z$; $K_z(0)$& The projected connected component on the considered projection plane; the $K_z$ containing the origin on the plane $\textbf{o}_z$ \\ \hline
    $\mathcal{F}$; $\mathcal{F}^{-1}$& The stereographic projection; the inverse stereographic projection\\ \hline
    $\theta(N,\gamma)$ & The percolation probability related to $N$ and $\gamma$\\ \hline
     \hline
    \end{tabular}
}
\end{center}
\label{tab:TableOfNotations}
%\vspace{-8mm}
\end{table*}

\section{Coverage Analysis} \label{sec:coverage}
\indent In this section, we first investigate the coverage analysis of LEO satellites. Then, we introduce how to project LEO satellite coverage areas on the sphere onto a tangent plane. Based on the stereographic projection, we define the percolation on the sphere and introduce the sub-critical and super-critical cases.

\subsection{Coverage Analysis of LEO Satellites}
\begin{figure}
    \centering
    \includegraphics[width=0.6\linewidth]{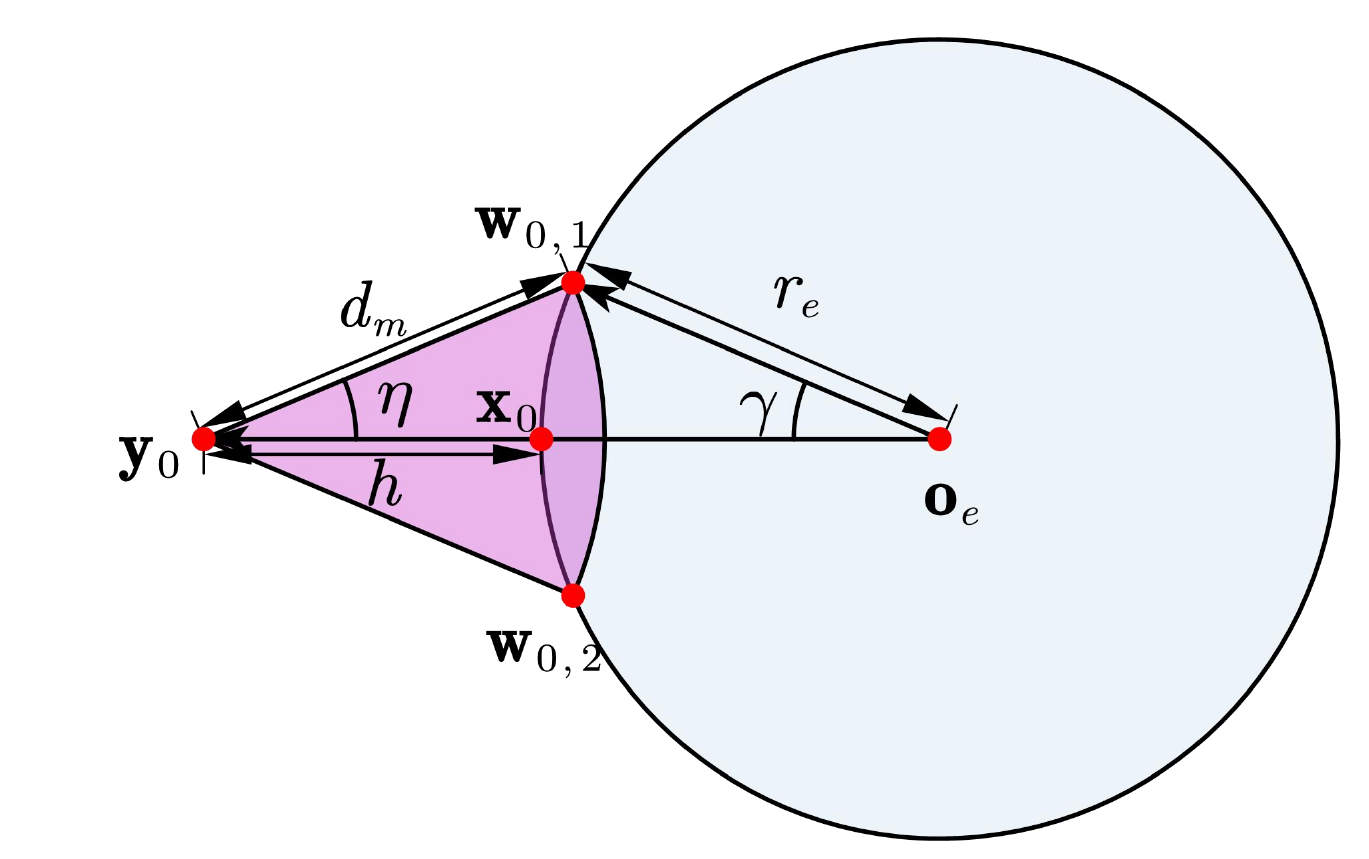}
    \caption{The geometric relationships between coverage angle $\gamma$, nadir angle $\eta$, satellite constellation altitude $h$ and maximum slant range $d_m$. }
    \label{fig:etagamma}
\end{figure}

\indent In Sec. \ref{sec:SysMod}, we introduce two different expressions of satellite coverage areas $\mathcal{A}_i=\mathcal{A}(\yi,\eta)$ and $\mathcal{O}_i=\mathcal{O}(\xi,\gamma)$. Because they represent the coverage area of the same LEO satellite, $\mathcal{O}(\xi,\gamma)=\mathcal{A}(\yi,\eta)$ must be satisfied. As shown in Fig.\ref{fig:etagamma}, we can obtain the relationships between $\gamma$, $\eta$, $h$ and $d_m$ as described in the below lemma.

\begin{lemma}\label{lem:etagamma}
   The relationships between the coverage angle $\gamma$, nadir angle $\eta$, constellation altitude $h$ and maximum slant range $d_m$ can be expressed as:
   \begin{equation}
       \gamma=\arcsin(\frac{d_m}{r_e}\sin\eta),
   \end{equation}
where
\begin{equation}
    d_m=-\sqrt{r_e^2-r_s^2\sin^2\eta}+r_s \cos\eta
\end{equation}
and
\begin{equation}
    0<\eta\leq \arcsin{\frac{r_e}{r_s}},\, r_s=r_e+h.
\end{equation}
\end{lemma}
\begin{IEEEproof}
    See Appendix~\ref{app:etagamma}.
\end{IEEEproof}
 Next, we introduce the coverage analysis of each point on the sphere in Theorem \ref{theo:covpro}.
\begin{theorem}\label{theo:covpro} Assume that the number of LEO satellites is $N$ and the coverage angle of each satellite is $
\gamma$. The probability of each point on the sphere being covered by at least one LEO satellites is:
    \begin{equation}
        p_{\rm cov}(N,\gamma)=1-\bigg(\frac{1+\cos\gamma}{2}\bigg)^{N}.
    \end{equation}
    \label{theo:pcov}
Correspondingly, the probability of each point on the sphere being not covered by any LEO satellite is:
    \begin{equation}
        p_{\rm ncov}(N,\gamma)=\bigg(\frac{1+\cos\gamma}{2}\bigg)^{N}.
    \end{equation}
    
\end{theorem}
\begin{IEEEproof}
    See Appendix~\ref{app:pcov}.
\end{IEEEproof}
It is worth noting that, for any point on the sphere, the sum of probabilities of being covered and being not covered equals 1, \ie $p_{\rm cov}+p_{\rm ncov}=1$. However, for any area $\mathcal{B}$ on the plane, we focus on the probability of it being completely covered or completely not covered to analyze the sub-critical case and super-critical case. Therefore, if we mention an event where $\mathcal{B}$ is `covered' or `not covered', they represent $\mathcal{B}$ is `completely covered' or `completely not covered'. These two probabilities must satisfy:  
\begin{equation}
    \P\{\mathcal{B} {\rm \; is\;covered}\}+\P\{\mathcal{B} {\rm \; is\;not\;covered}\}\leq 1
\end{equation}
because $\P\{\mathcal{B} {\rm \; is\;partially\;covered}\}\geq 0$.

\subsection{Percolation through Stereographic Projection}
\begin{figure}[ht]
    \centering
    \includegraphics[width=0.75\linewidth]{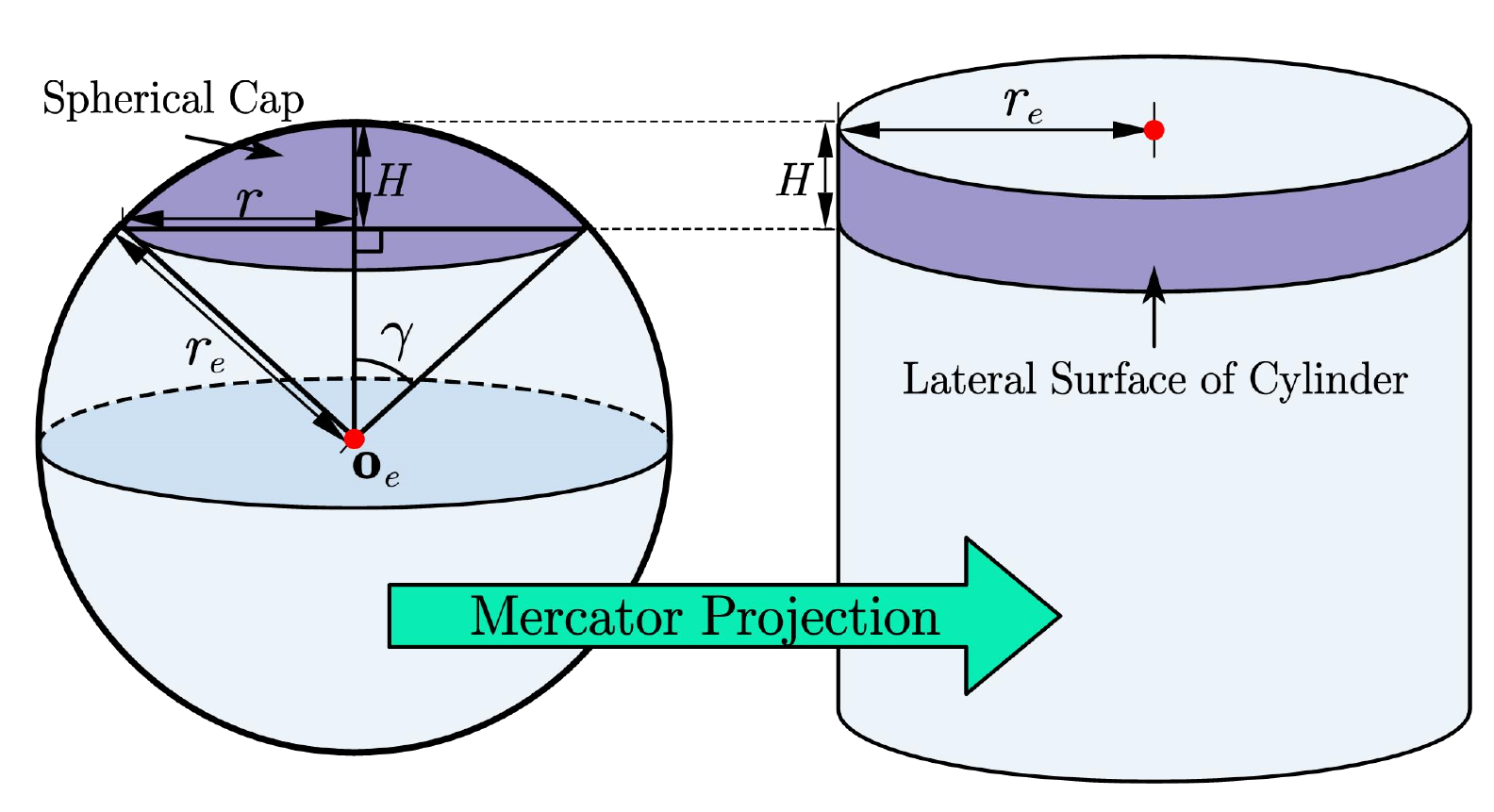}
    \caption{The area of spherical cap and the Mercator projection. }
    \label{fig:caparea}
\end{figure}

\indent To analyze the percolation on the sphere, we need to separate the whole sphere using some special lattice. Classical percolation analysis is always based on triangular, square or hexagonal lattices. Unlike a plane, the sphere can not be divided using a homogeneous lattice. Mercator projection in Fig.\ref{fig:caparea} can map the sphere on a square area, which is commonly used in geography \cite{snyder1987map}, however, the spherical coverage areas of LEO satellites become irregular and not tractable. Therefore, we propose to use the stereographic projection to analyze the percolation on the sphere \cite{snyder1987map}, which is a specific example of Alexandroff extension mapping a sphere onto a plane \cite{willard2012general}. The stereographic projection is introduced in Theorem \ref{theo:stereo}.

\begin{figure}
    \centering
    \includegraphics[width=0.65\linewidth]{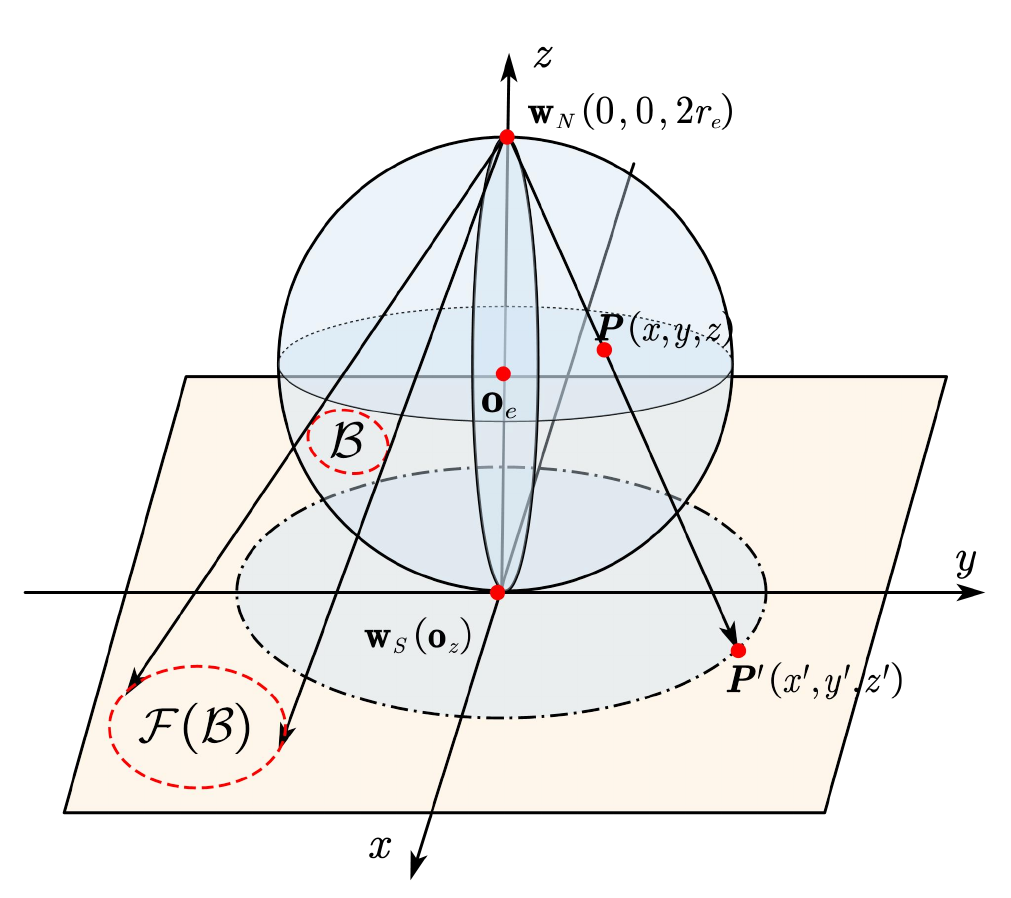}
    \caption{The stereographic projection. The projection plane is on the $xo_zy$ plane, which is tangent to the earth on the South Pole $\textbf{w}_S(\textbf{o}_z)$. The earth's center $\textbf{o}_e$ and North Pole $\textbf{w}_N$ are both on the z-axis. $P'$ is the stereographic projection of $P$. Any circle on the sphere corresponds to a circle on the projection plane. If the spherical cap excludes the North Pole $\textbf{w}_N$, the spherical cap is projected to a finite circular area. Inversely, any finite circular area corresponds to a spherical cap excluding $\textbf{w}_N$.}
    \label{fig:stereographic}
\end{figure}

\begin{theorem}
\label{theo:stereo}
    As shown in Fig.\ref{fig:stereographic}, $P(x,y,z)$ is a point on the sphere and $P'(x',y',z')$ is its stereographic projection on the projection plane. The stereographic projection $P'=\mathcal{F}(P)$ leads to:
\begin{equation}
\begin{array}{r@{}l}
(x',y',z')
=\displaystyle\bigg(\frac{2r_e}{2r_e-z}x,\frac{2r_e}{2r_e-z}y,0\bigg),
\end{array}
\end{equation}
and the inverse stereographic projection $P=\mathcal{F}^{-1}(P')$ leads to:
\begin{equation}
\begin{array}{l}
    (x,y,z)\\=\displaystyle\bigg(\frac{4r_e^2 x'}{4r_e^2+x'^2+y'^2},\frac{4r_e^2 x'}{4r_e^2+x'^2+y'^2},\frac{2r_e(x'^2+y'^2)}{4r_e^2+x'^2+y'^2}\bigg).
\end{array}
\end{equation}
\end{theorem}
\begin{IEEEproof}
    Notice that the South Pole of the sphere is the origin of the projection plane, \ie $\textbf{w}_S=\textbf{o}_{z}$. The coordinate relationship in the stereographic projection and inverse stereographic projection can be easily proved using $\overrightarrow{w_{N}P}=\frac{2r_e-z}{2r_e}\overrightarrow{w_{N}P'}$.  
\end{IEEEproof}
Except for the North Pole $w_N$, the stereographic projection is a bijection between a sphere and a plane. Therefore, we introduce a property of stereographic projection in Lemma \ref{lem:subset}.

\begin{lemma}\label{lem:subset}
    Define the mapping from the sphere to the plane through stereographic projection as a function $\mathcal{F}$. For two spherical areas on the sphere $\mathcal{B}$ and $\mathcal{C}$, their projections on the plane satisfy:
    \begin{equation}
        \mathcal{B}\subseteq \mathcal{C} \Leftrightarrow \mathcal{F}(\mathcal{B})\subseteq \mathcal{F}(\mathcal{C}).
        \label{mappingrelation}
    \end{equation}
    \label{lem:mappingrelation}
\end{lemma}

\begin{IEEEproof}
    As a kind of bijection, stereographic projection, except for the North Pole  $\textbf{w}_N$, has the same property as bijection.
\end{IEEEproof}
\begin{remark}
    Notice that a closed shape on the sphere is not always projected to a closed shape on the projection plane. Once the North Pole is included in the $\mathcal{B}$, the size of projection $\mathcal{F}(\mathcal{B})$ goes to infinite. However, the property (\ref{mappingrelation}) still holds.
\end{remark}

% Based on Lemma \ref{lem:subset}, we can show the below corollary:
% \begin{corollary}
%     If a spherical area $\mathcal{B}$ is covered by a LEO satellite's coverage area $\mathcal{A}_i$, \ie $\mathcal{B}\subseteq\mathcal{A}_i$, their projections on the plane also satisfy $\mathcal{F}(\mathcal{B})\subseteq\mathcal{F}(\mathcal{A}_i)$, vice versa, that is:
%     \begin{equation}
%         \mathcal{B}\subseteq \mathcal{A}_i \Leftrightarrow \mathcal{F}(\mathcal{B})\subseteq \mathcal{F}(\mathcal{A}_i).
%         \label{corsubset2}
%     \end{equation}
% \label{cor:mapspherical}
% \end{corollary}
\begin{figure}
    \centering
    \includegraphics[width=0.8\linewidth]{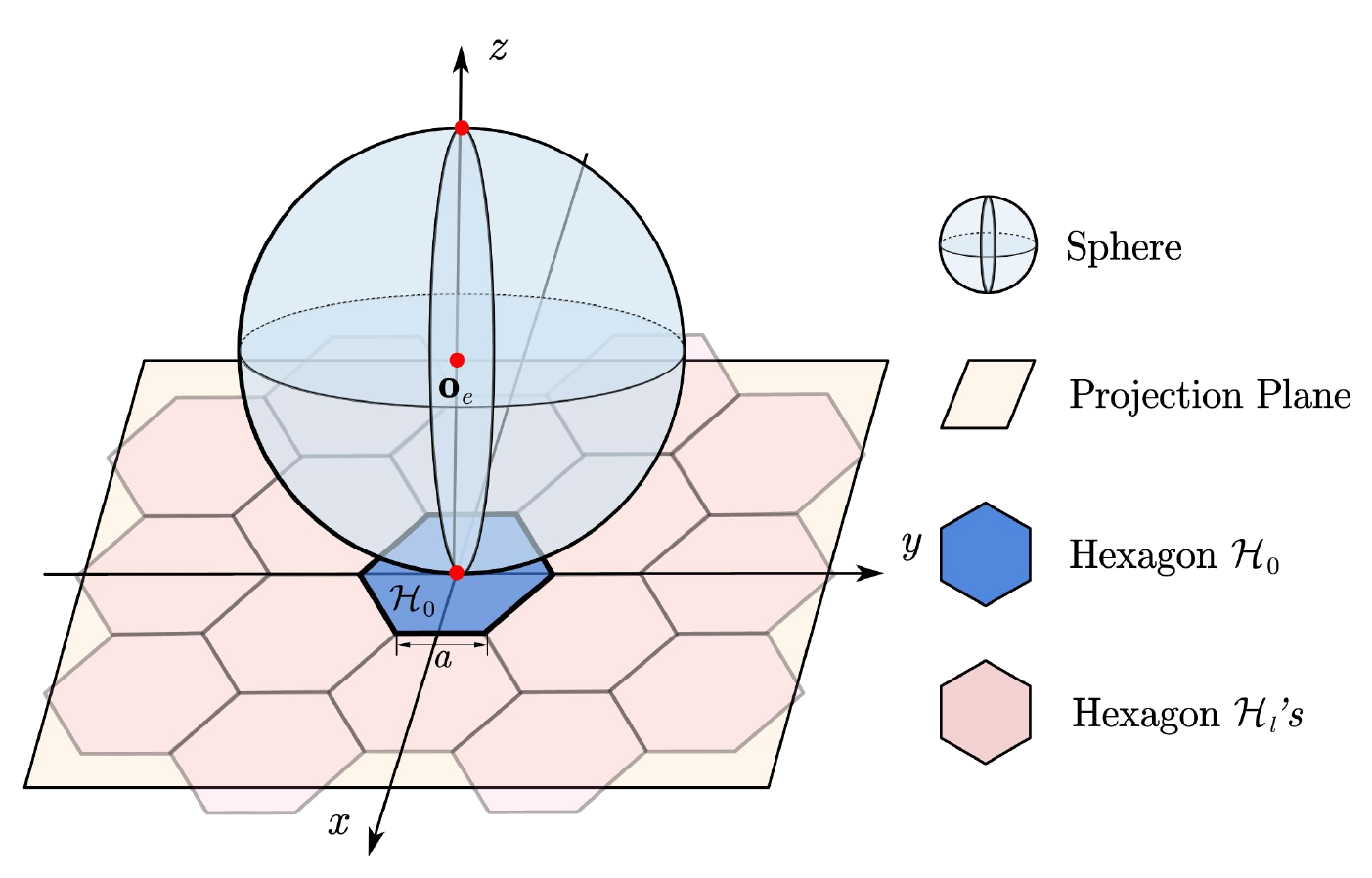}
    \caption{Hexagonal lattice on the projected plane. The side length of hexagons is $a$ and $\mathcal{H}_0$ is the hexagon which is centered at the origin.}
    \label{fig:hexagon}
\end{figure}
\indent To make the percolation on the sphere a tractable problem, we propose to discuss the percolation on the stereographic projection plane. As shown in Fig.\ref{fig:hexagon}, we define the homogeneous hexagons on the plane as $\mathcal{H}_l$'s with the side length $a$. Through inverse stereographic projection, we can also find the original area of $H_l$ on the sphere, \ie $\mathcal{F}^{-1}(\mathcal{H}_l)$. For percolation on the sphere, we focus on whether $\mathcal{F}^{-1}(\mathcal{H}_l)$ can be covered by the LEO satellites' coverage areas, \ie $\mathcal{A}_i$'s. Using the property (\ref{mappingrelation}), the problem is equivalent to whether the hexagon $\mathcal{H}_l$ on the plane can be covered by the projections of $\mathcal{A}_i$'s, \ie $\mathcal{F}(\mathcal{A}_i)$'s.
% \begin{corollary}
%     If a hexagon on the plane $\mathcal{H}_l$ is covered by the projection of a LEO satellite's coverage area $\mathcal{F}(\mathcal{A}_i)$, \ie $\mathcal{H}_l\subseteq \mathcal{F}(\mathcal{A}_i)$, their original areas on the sphere also satisfy $\mathcal{F}^{-1}(\mathcal{H}_l)\subseteq\mathcal{A}_i$, vice versa, that is:
%     \begin{equation}
%         \mathcal{H}_l\subseteq \mathcal{F}(\mathcal{A}_i) \Leftrightarrow \mathcal{F}^{-1}(\mathcal{H}_l)\subseteq\mathcal{A}_i.
%         \label{corsubset}
%     \end{equation}
% \label{cor:maphexagon}
% \end{corollary}
 On a plane, the face percolation of hexagons means there exist giant components whose cardinality is infinite. As shown in (\ref{defineper}), percolation probability is always defined as the probability of generating a giant component that crosses the origin. Through inverse stereographic projection, such a giant 
 component is projected from a continuous coverage area from the South Pole to the North Pole. This also represents the `farthest coverage on the earth'. Therefore, we propose to define the percolation probability on the sphere as below.
\begin{definition}
    On a sphere, percolation probability is defined as the probability of generating continuous coverage areas that contain two symmetry points about the sphere's center. Especially, we can also define it using the probability of connecting the North Pole and the South Pole of earth, \ie
\begin{equation}
    \theta(N,\gamma)=\P\{\textbf{w}_N\in K_x(\textbf{w}_S)\}.
\end{equation} 
where $K_x$ denotes the giant component on the sphere, $\textbf{w}_S$ and $\textbf{w}_N$ represent the South Pole and the North Pole, respectively. $K_x(\textbf{w}_S)$ represents the connected component starting from the South Pole.
\label{def:concomsphere}
\end{definition}

\begin{remark}
    On the earth, the spherical distance between any two points is less than or equal to $\pi r_e$, that is, the two points that are symmetric about the earth's center have the maximum spherical distance. Unlike the analysis on an infinite plane, the cardinality of the connected component will not reach infinity. The farthest spherical distance it can reach is determined, which can be used as a judgement of percolation. In addition, the cardinality of the connected component has its upper bound, that is, the entire sphere.
\end{remark}
\indent As a basic knowledge of hexagonal face percolation, the sufficient and necessary condition for face percolation of hexagons is that the probability of each hexagon being covered should be larger than $\frac{1}{2}$, \ie
\begin{equation}
    \theta(N,\gamma)>0,\,{\rm if}\,\P\{\mathcal{H}_l\,{\rm is\, covered}\}>\frac{1}{2}.
\end{equation}
\indent It is difficult to calculate $\P\{\mathcal{H}_l\,{\rm is\, covered}\}$ directly because the shape of $\mathcal{F}^{-1}(\mathcal{H}_l)$ is irregular. However, we can use some circular areas to help find the tight upper bound and lower bound of it. At the same time, the coverage areas of LEO satellites are typically modeled as circular areas. Therefore, we introduce the below lemma.
\begin{lemma}
    For a spherical cap on the sphere, its projection on the plane is a circular area, unless the border of the spherical cap crosses the top of the sphere.
    Inversely, for each circular area on the projected plane, its inverse projection on the sphere is a circular area. 
    \label{lem:captocircle}
\end{lemma}
\begin{IEEEproof}
    See the proof in \cite[88.1]{pedoe2013geometry}. %See Appendix~\ref{app:captocircle}.
\end{IEEEproof}
\begin{remark}
    For stereographic projection, the North Pole $\textbf{w}_N$ is considered the top of the earth. There exist three cases: \textbf{i) the spherical cap excludes the top}, where the projection is a closed circular area on the plane, \textbf{ii) the border of the spherical cap crosses the top}, where the projection is a region divided by a straight line that does not include the origin $\textbf{o}_z$, \textbf{iii) the spherical cap includes the top}, where the projection is open and its inner envelope is a circular area on the plane. On the other hand, for inverse stereographic projection, closed circular areas on the plane can be always projected to spherical caps on the sphere, which does not contain the North Pole $\textbf{w}_N$.
\end{remark}
% According to Lemma \ref{lem:captocircle}, we introduce the below corollary:
% \begin{corollary}
%     For a LEO satellite's coverage area $\mathcal{O}(\xi,\gamma)=\mathcal{A}(\yi,\eta)$, its projection on the plane is a circular area unless the angle $\angle \textbf{w}_{N}\textbf{o}_e\yi=\gamma$ where $\textbf{w}_{N}$ is the top of the sphere.
% \end{corollary}
% \begin{IEEEproof}
%     Let the coverage area $\mathcal{O}(\xi,\gamma)$ be the spherical cap in Lemma \ref{lem:captocircle}, its coverage is a circular area on the plane. When $\angle \textbf{w}_{N}\textbf{o}_e\yi=\gamma$, the considered plane crosses the top of the sphere, and the projection of the LEO satellite's coverage area is a straight line on the plane.
% \end{IEEEproof}
% \indent Therefore, almost all LEO satellites' coverage areas on the sphere are projected to circular areas on the plane. We can obtain the radius of these coverage areas. Next, we introduce Lemma \ref{lem:circularradius}:
\indent In this paper, we need to first analyze the property of circular areas on the projection plane. As introduced in Lemma \ref{lem:captocircle}, their inverse projections on the sphere can be always modeled as circular areas that does not contain the North Pole $\textbf{w}_N$. So that, we introduce the relationship between central angle of a spherical cap and the radius of its projected circular area.

\begin{lemma}\label{lem:circularradius}
     For a spherical cap that does not contain the North Pole $\textbf{w}_N$ with a central angle $\gamma_0$, the radius of its projected circular area on the projection plane can be expressed as:
\begin{equation}
    r(\psi)=r_e\bigg|\tan(\frac{\psi+\gamma_0}{2})-\tan(\frac{\psi-\gamma_0}{2})\bigg|        
\label{radiusprojection}
\end{equation}
where $\psi=\angle \textbf{w}_S \oe \textbf{x}<\pi-\gamma_0$ and $\textbf{x}$ is the center of the spherical cap. The range of $r(\psi)$ is:
\begin{equation}
    2r_e\tan(\frac{\gamma}{2})\leq r(\psi)<+\infty.
\end{equation}
Conversely, for a circular area on the projection plane with a radius $r$, the central angle of its original spherical cap is upper and lower bounded as follows:
\begin{equation}
    0<\gamma_0\leq 2\arctan(\frac{r}{2r_e}).
\end{equation}
\end{lemma}
\begin{IEEEproof}
    See Appendix~\ref{app:circularradius}.
\end{IEEEproof}
\indent Lemma \ref{lem:captocircle} and Lemma \ref{lem:circularradius} already exhibit how to project the spherical caps on the sphere to its tangent projection plane, which are used to do the critical analysis in the next section.

\section{Critical Analysis}\label{sec:critical}
In this section, we first prove that percolation probability is a non-decreasing function of the number of satellites. Next, We introduce the sub-critical and super-critical cases where the percolation probability is zero and non-zero, respectively. Based on these, we prove the existence of the critical number of LEO satellites to realize the phase transition of percolation probability on the sphere. After that, we discuss the critical case and derive a closed-form expression of the critical satellite number $N_c$.  
\subsection{Phase Transition}\label{subsec:phasetransition}
  
 \indent In order to prove the existence of phase transition of percolation probability and derive the critical number of satellites, we first introduce the relationship between the percolation probability $\theta$ and the number of satellites $N$ in the below lemma.
\begin{lemma}\label{lem:nondecreasing}
    When the LEO satellites are deployed at the same altitude following a BPP with a coverage angle $\gamma$, the percolation probability and the number of LEO satellites satisfy:
    \begin{equation}
        \theta(N_1,\gamma)\leq\theta(N_2,\gamma), \;{\rm for}\; 0<N_1<N_2,
    \end{equation}
when the value of $\gamma$ is fixed.
\end{lemma}
\begin{IEEEproof}
    See Appendix~\ref{app:nondecreasing}.
\end{IEEEproof}
\indent Therefore, the percolation probability $\theta(N,\gamma)$ does not decrease as $N$ increases. Next, we introduce a sub-critical case to obtain a lower bound $N_L$ of the critical number of LEO satellites, where percolation probability is zero when $N\leq N_L$.\\

\begin{figure}
    \centering
    \includegraphics[width=0.4\linewidth]{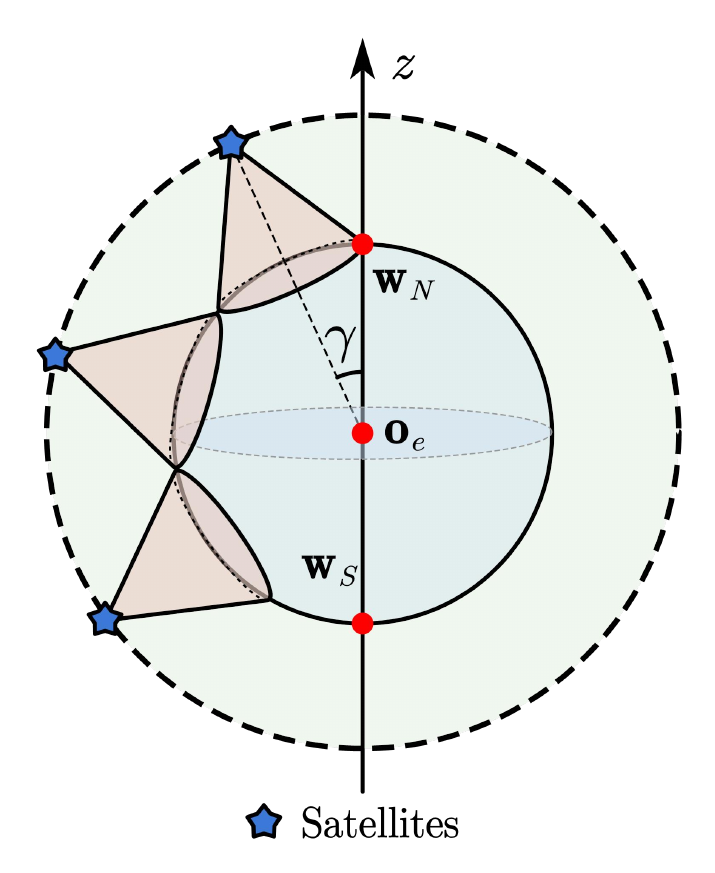}
    \caption{Sub-critical case: All satellites are deployed on the same meridian, where neighbour coverage areas are tangent to each other. However, the longest spherical distance inside the coverage areas does not exceed $\pi r_e$ when the number of satellites is not large enough.}
    \label{fig:lowerbound}
\end{figure}

\noindent \textbf{Sub-critical case:} We choose a certain meridian (e.g. the prime meridian). The LEO satellites are deployed along the longitude and the borders of two adjacent coverage areas are tangent. If the longest spherical distance inside the covered areas is less than $\pi r_e$, the probability of percolation is 0. Therefore, we first introduce the sufficient condition for zero percolation probability in the below lemma. 
\begin{lemma}\label{lem:lowerbound}
    When the number of LEO satellites is less than $N_L$, the percolation probability $\theta(N,\gamma)$ is zero, \ie
\begin{equation}
    \theta(N,\gamma)=0\;{\rm if}\;N\leq N_L.
\end{equation}    
    The expression of $N_L$ is
\begin{equation}
    N_L = \left \lfloor \frac{\pi}{2\gamma} \right \rfloor 
\label{lowerbound}
\end{equation}
where $\left \lfloor x \right \rfloor$ denotes the largest integer less than x, and $\gamma$ is the coverage angle of each LEO satellite.
\end{lemma}
\begin{IEEEproof}
    As shown in Fig.\ref{fig:lowerbound}, when $N\leq\left \lfloor \frac{\pi}{2\gamma} \right \rfloor$, even though the satellites are deployed on the same orbit, any continuous coverage path containing $\textbf{w}_N$ and $\textbf{w}_S$ can not be generated.
\end{IEEEproof}

\begin{corollary}
    If the critical number of satellites $N_c$ for the phase transition of percolation probability exists, $N_L$ is the lower bound of $N_c$, \ie $N_c\geq N_L$.
    \label{cor:subcritical}
\end{corollary}

Next, in order to obtain an upper bound of the critical number of LEO satellites, where percolation probability is non-zero when $N\geq N_U$, we need to ensure that the percolation probability has a computable and non-zero lower bound when $N=N_U$. Therefore, we introduce a super-critical case as shown below.\\ 

\begin{figure}
    \centering
    \includegraphics[width=1\linewidth]{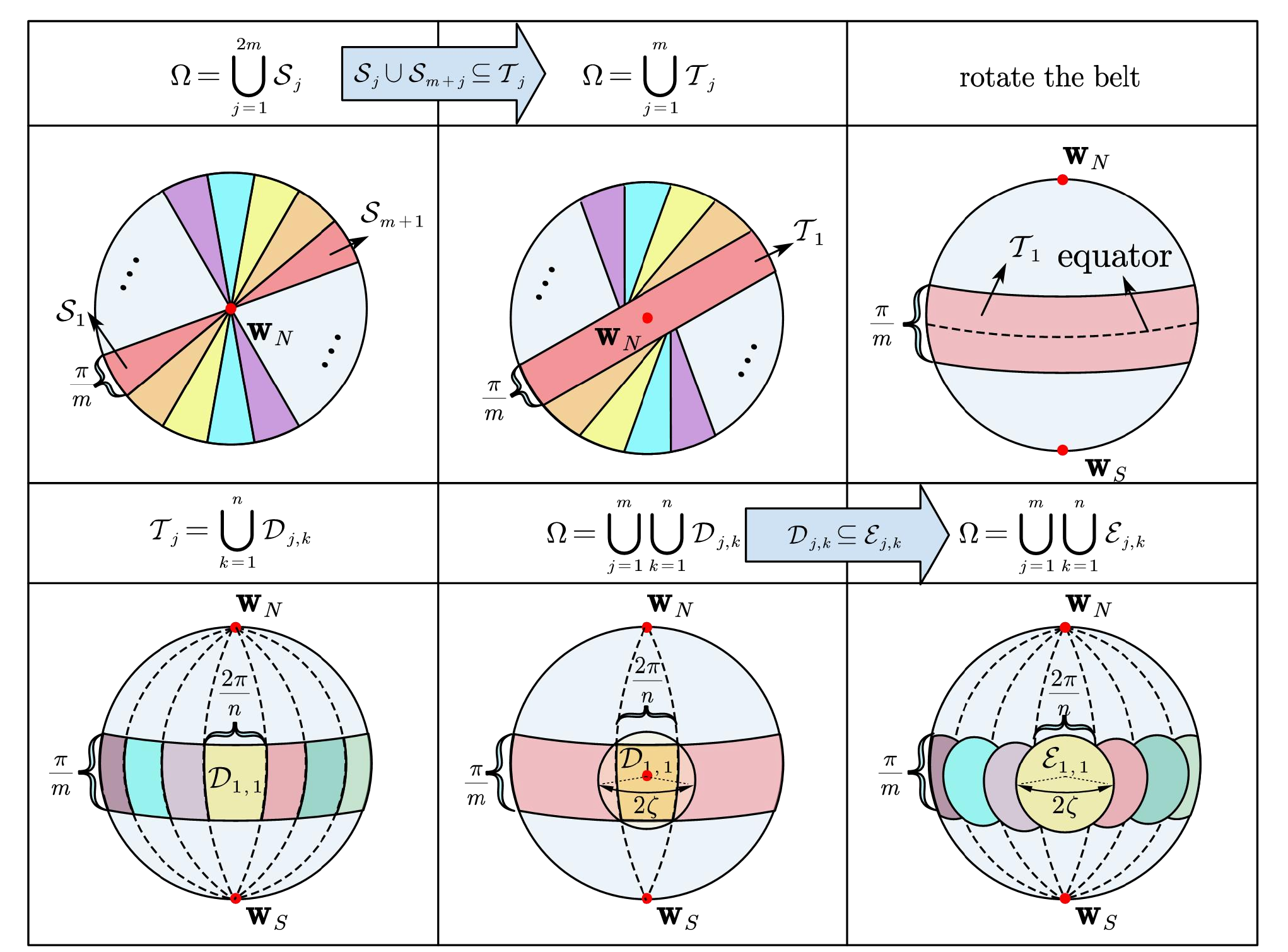}
    \caption{Super-critical case: a full coverage scheme. Above (from left to right): \textit{a)} The whole sphere is firstly divided into $2m$ slices. \textit{b)} Because each belt can contain two slices that are symmetric, the whole sphere can be considered as the union of $m$ belts. \textit{c)} Rotate the belt. Below (from left to right): \textit{d)} Each belt can be divided into $n$ pieces. \textit{e)} The whole sphere can be consider the union of $m\times n$ pieces. \textit{f)} Because each piece can be contained by a spherical cap, the whole sphere can be considered the union of $m\times n$ spherical caps.}
    \label{fig:Upperbound}
\end{figure}

\noindent \textbf{Super-critical case:} This super-critical case is designed in six steps: \textit{i)} Along the meridians, we divide the whole sphere into $2m$ `slices', where each slice spans $\frac{\pi}{m}$ of longitude. \textit{ii)} Each two slices symmetric about the earth's center can be contained by a `belt'. Therefore, $m$ of belts can cover the whole sphere. \textit{iii)} Rotate a belt and make it symmetric about the equatorial plane, it can be considered a belt spanning $\frac{\pi}{m}$ of latitude. \textit{iv)} By dividing such a belt into $n$ uniform `pieces', we can use in total $m\times n$ pieces to cover the whole sphere. Each piece spans $\frac{\pi}{m}$ of longitude and $\frac{2\pi}{n}$ of latitude. \textit{v)} Each piece can be contained by a spherical cap which is smaller than the coverage area of a satellite. \textit{vi)} Use the $m\times n$ of satellites to cover the target spherical caps one by one. The steps from i) to v) are shown in Fig.\ref{fig:Upperbound}, which explains how to represent the whole sphere using the union of spherical caps. 
\\
\indent In the super-critical case, we need to design $m$ and $n$ large enough to make such a full coverage deployment feasible and obtain a computable non-zero lower bound of percolation probability. Therefore, we introduce the sufficient condition for non-zero percolation probability in the below lemma.
\begin{lemma}\label{lem:upperbound1}
    When the number of LEO satellites is larger than $N_U$, the percolation probability $\theta(N,\gamma)$ is non-zero, \ie
\begin{equation}
    \theta(N,\gamma)>0\;{\rm if}\;N\geq N_U. 
\end{equation}    
    The expression of $N_U$ is 
\begin{equation}
    N_U = m\times n 
\label{upperbound}
\end{equation}
with
\begin{equation}
    m = \left \lceil \frac{\pi}{\gamma} \right \rceil,\;n= \left \lceil \frac{\pi}{\arccos{\frac{\cos\gamma}{\cos\frac{\pi}{2m}}}} \right \rceil +1
\label{upperboundmn}
\end{equation}
where $\left \lceil x \right \rceil$ denotes the smallest integer greater than x and $\gamma$ is the coverage angle of each LEO satellite.
\end{lemma}
\begin{IEEEproof}
    See Appendix~\ref{app:upperbound}. 
\end{IEEEproof}
\begin{corollary}
    If the critical number of satellites $N_c$ exists, $N_U$ is the upper bound of $N_c$, \ie $N_c\leq N_U$.
    \label{cor:supercritical}
\end{corollary}
% \begin{corollary}
%     The upper bound in Lemma \ref{lem:upperbound1} shows that: if the number of LEO satellites is greater than $N_U$, \ie $N>N_L$ the percolation probability $\theta(N,\gamma)>0$.
%     \label{cor:supercritical}
% \end{corollary}
% \begin{remark}
%     When the number of satellites tends to $\infty$, the coverage probability of each point on the sphere $p_{cov}$ is almost surely 1. In this case, the percolation probability is almost surely 1, \ie $\theta(\infty,\gamma)=1$, almost surely. This extreme case can be used to further prove that the existence of upper bound of the critical number of satellites.
%     \label{rem:upperboundinfty}
% \end{remark}

 Based on the sub-critical and super-critical cases, we can prove the existence of the critical value of $N$, \ie $N_c$, in the following lemma.

\begin{lemma}\label{lem:phasetransition}
    When the LEO satellites are deployed at the same altitude following a BPP with a fixed value of $\gamma$, there exists a critical value $N_c$ satisfying:
    \begin{equation}
    \begin{array}{c}
        \theta(N,\gamma)=0,\, {\rm for}\; N<  N_c,\\
        \theta(N,\gamma)>0,\, {\rm for}\; N> N_c. 
    \end{array}
    \end{equation}
where $N_L\leq \left\lfloor N_c\right\rfloor$ and $\left\lceil N_c\right\rceil\leq N_U$.
\end{lemma}
\begin{IEEEproof}
    See Appendix~\ref{app:phasetransition}.
\end{IEEEproof}
Therefore, we prove the existence of the critical value of $N$, \ie $N_c$, which exhibits the phase transition of percolation probability.
\subsection{Tight bounds and critical analysis}
 In Lemma \ref{lem:phasetransition}, we have proved that the critical number of LEO satellites for phase transition of percolation probability exists. In this part, we propose to use the stereographic projection to find a tight lower bound and a tight upper bound for $N_c$, and introduce the closed-form expression of $N_c$.\\
 
 \noindent\textbf{Hexagonal face percolation on the projection plane:} As shown in Definition \ref{def:concomsphere}, the percolation on the sphere containing the South Pole $\textbf{w}_S$ represents the percolation on the projection plane containing the origin $\textbf{o}_z$. We first consider the hexagons with side length $a$. In percolation theory, if all hexagonal faces have the same probabilities $\P\{\mathcal{H}_{l} {\rm \; is\;open}\}$ and $\P\{\mathcal{H}_{l} {\rm \; is\;closed}\}$, we have: \textit{i) $\theta=0$ when $\P\{\mathcal{H}_{l} {\rm \; is\;closed}\}>1/2$ }and\textit{ ii) $\theta>0$ when $\P\{\mathcal{H}_{l} {\rm \; is\;open}\}>1/2$}. To find the tight bounds, we first introduce the below theorem.

 \begin{figure}
    \centering
    \includegraphics[width=0.6\linewidth]{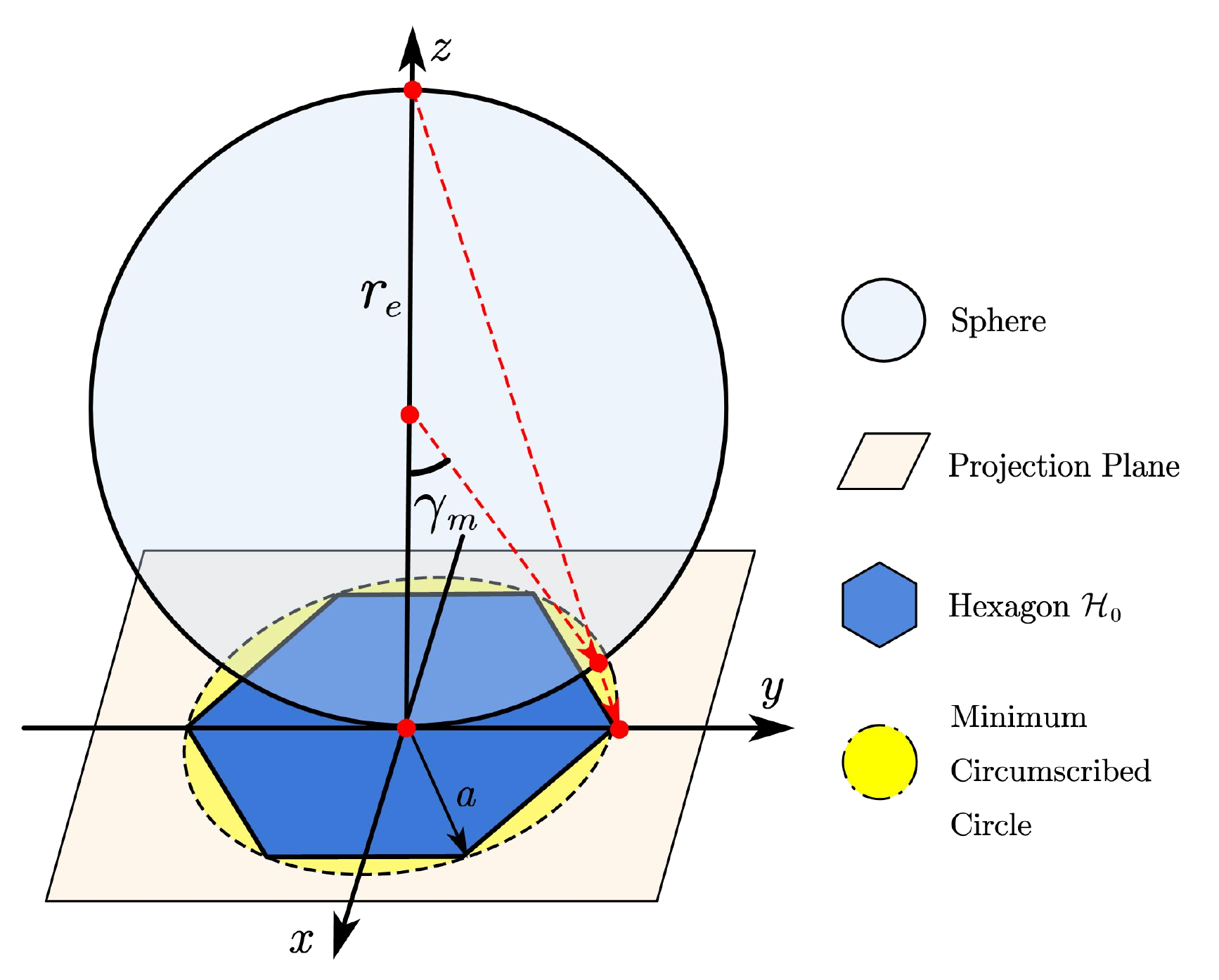}
    \caption{The minimum circumscribed circle of the hexagon on the projected plane, which is centered at the origin of the projection plane. Its central angle of the corresponding original spherical cap is the maximum one, that is $\gamma_m$. }
    \label{fig:gamma}
\end{figure}

\begin{theorem}
    Consider the hexagonal lattice on the plane where the side length of each hexagon is $a$. If the coverage probabilities of different hexagons are different,
    the sufficient condition for non-zero face percolation probability is:
\begin{equation}
    \P\{\mathcal{H}_{l} {\rm \; is\;open}\}>1/2,
\label{sufficientcover}
\end{equation}
and the sufficient condition for zero face percolation probability is:
\begin{equation}
    \P\{\mathcal{H}_{l} {\rm \; is\;closed}\}>1/2.
\label{sufficientnotcover}
\end{equation}
\label{theo:inhomohexagon}
\end{theorem}
\begin{IEEEproof}
    See Appendix~\ref{app:inhomohexagon}.
\end{IEEEproof}
\indent Next, we introduce the lower bounds $\P\{\mathcal{H}_{l} {\rm \; is\;open}\}$ and $\P\{\mathcal{H}_{l} {\rm \; is\;closed}\}$ in the below lemma.

\begin{lemma}\label{lem:boundsforhexagons}
Let $\P\{\mathcal{H}_{l} {\rm \; is\;open}\}$ denote the probability of a hexagon $\mathcal{H}_{l}$ on the projection plane being covered by LEO satellites. 
The lower bound of $\P\{\mathcal{H}_{l} {\rm \; is\;open}\}$ is shown as:
\begin{equation}
\begin{array}{c}
    \P\{\mathcal{H}_{l} {\rm \; is\;open}\}\geq\displaystyle 1-\bigg(\frac{1+\cos(\gamma-\gamma_m)}{2}\bigg)^{N}, 
\end{array}
\label{mincoverageprob}
\end{equation}
where
\begin{equation}
    \gamma_m=2\arctan \frac{a}{2 r_e}.
\end{equation}
Let $\P\{\mathcal{H}_{l} {\rm \; is\;closed}\}$ denote the probability of the hexagon $\mathcal{H}_{l}$ on the projection plane being not covered by LEO satellites. The lower bound of $\P\{\mathcal{H}_{l} {\rm \; is\;closed}\}$ is shown as:
\begin{equation}
\begin{array}{c}
    \P\{\mathcal{H}_{l} {\rm \; is\;closed}\}\geq\displaystyle\bigg(\frac{1+\cos(\gamma+\gamma_m)}{2}\bigg)^{N}.
\end{array}
\label{minnotcoverageprob}
\end{equation}
% where
% \begin{equation}
%   \gamma_m=2\arctan \frac{a}{r_e}.
% \end{equation}
\label{lem:minprobs}
\end{lemma}
\begin{IEEEproof}
    See Appendix~\ref{app:boundsforhexagons}.
\end{IEEEproof}

Substitute (\ref{mincoverageprob}) and (\ref{minnotcoverageprob}) in Lemma \ref{lem:minprobs} into the sufficient conditions for non-zero or zero percolation probability (\ref{sufficientcover}) and (\ref{sufficientnotcover}) in Theorem \ref{theo:inhomohexagon}, we can obtain the sufficient conditions of the number of LEO satellites for non-zero or zero percolation probability that are shown in the below theorem.
\begin{theorem}
Given that the coverage angle of each satellite is $\gamma$, $r_e$ is the radius of the earth and $a$ is the side length of hexagons on the projection plane. The sufficient condition of the number of LEO satellites for non-zero percolation probability is:
\begin{equation}
    N>N_c^U
\end{equation}
where 
\begin{equation}
    N_c^U=\frac{\ln 2}{\ln 2-\ln(1+\cos(\gamma-2\arctan \frac{a}{2r_e}))}
\end{equation}

\noindent is the upper bound of critical number of LEO satellites for phase transition of percolation probability, \ie $N_c\leq N_c^U$.\\
\indent The sufficient condition of the number of LEO satellites for zero percolation probability is:
\begin{equation}
    N<N_c^L
\end{equation}
where 
\begin{equation}
    N_c^L=\frac{\ln 2}{\ln 2-\ln(1+\cos(\gamma+2\arctan \frac{a}{2r_e}))}
\end{equation}
is the lower bound of critical number of LEO satellites for phase transition of percolation probability, \ie $N_c\geq N_c^L$.
\label{theo:sufficientconditions}
\end{theorem}
\begin{IEEEproof}
The upper bound $N_c^{U}$ and lower bound $N_c^{L}$ are obtained by substituting (\ref{mincoverageprob}) and (\ref{minnotcoverageprob}) in Lemma \ref{lem:minprobs} into the sufficient conditions for non-zero or zero percolation probability (\ref{sufficientcover}) and (\ref{sufficientnotcover}) in Theorem \ref{theo:inhomohexagon}, respectively.
\end{IEEEproof}
\indent To analyze the continuous percolation on the sphere, we also consider the continuous percolation on the plane. Therefore, the side length of considered hexagons on the plane is assumed to approach 0. We obtain the explicit expression for the critical number of LEO satellites in the below lemma.
\begin{lemma}
The critical number of LEO satellites for phase transition of percolation probability is:
\begin{equation}
N_c=\frac{\ln 2}{\ln 2-\ln(1+\cos\gamma)}.
\label{criticalNc}
\end{equation}
\label{lem:criticalanalysis}
\end{lemma}
\begin{IEEEproof}
See Appendix~\ref{app:criticalanalysis}. 
\end{IEEEproof}
% \begin{remark}
% We obtain the lower bound $N_L$ and upper bound $N_U$ through coverage analysis on the sphere. We also obtain the tight bounds $N_c^L$ and $N_c^U$ through percolation on the considered projection plane, and further obtain the critical number of LEO satellites for phase transition of percolation probability, \ie $N_c$. It is necessary to verify the relationship between the $N_L$, $N_U$ and $N_c$, which is also shown in the proof of Lemma \ref{lem:criticalanalysis}.
% \end{remark}
$N_c$ is the only explicit value which is always located between bounds $N_c^{L}$ and $N_c^{U}$. At the same time, the upper and lower bounds $N_c^{L}$ and $N_c^{U}$ are both tighter than $N_L$ and $N_U$ when $a$ approaches 0. Theoretically, for $N\geq\left \lceil N_c \right \rceil$, $\theta(N,\gamma)>0$ and for $N\leq\left \lfloor N_c \right \rfloor$, $\theta(N,\gamma)=0$.\\
\indent It is worth noting that, the critical number $N_c$ is the same as the solution of $p_{\rm cov}(N,\gamma)=1/2$ or $p_{\rm ncov}(N,\gamma)=1/2$. When $N>N_c$, $ p_{\rm cov}(N,\gamma)>1/2$. When $N<N_c$, $p_{\rm ncov}(N,\gamma)>1/2$ and $p_{\rm cov}(N,\gamma)<1/2$. Therefore, we obtain the critical condition for phase transition of percolation probability in the below theorem.
\begin{theorem}
    Assume that all points on the sphere has the same probability of being covered, that is $p_{\rm cov}$. 
    The phase transition of continuous percolation on the sphere is expressed as:
\begin{equation}
\begin{array}{r@{}l}
    \theta(p_{\rm cov})=0,\;& for\;p_{\rm cov}<\frac{1}{2}, \\\theta(p_{\rm cov})>0,\;& for\;p_{\rm cov}>\frac{1}{2}. \\
    
\end{array}
\end{equation}
where $p_{\rm cov}$ represents the homogenerous coverage probability on the sphere and $\theta(p_{\rm cov})$ is the percolation probability based on this coverage probability.
\label{theo:perpro_covpro}
\end{theorem}
In this paper, the coverage probability and the percolation probability both depend on the number of LEO satellites $N$ and its coverage angle $\gamma$. We have used the expression $\theta(N,\gamma)$ for percolation probability. Similar to Lemma \ref{lem:criticalanalysis}, for a fixed number of LEO satellites, the relationship between the critical constellation altitude $h^c$ and maximum slant range $d_m$ is shown in the following lemma.
\begin{lemma}
    When the number of LEO satellites is fixed, the critical constellation altitude $h$ can be expressed using the maximum slant range $d_m$:
    \begin{equation}
        h^c = \sqrt{d_m^2-r_e^2+t^2(N) r_e^2}+t(N)r_e-r_e.
        \label{criticalaltitude}
    \end{equation}
    Correspondingly, the critical maximum slant range $d_m^c$ can be also expressed using the constellation altitude $h$:
    \begin{equation}
        d_m^c=\sqrt{r_e^2+(r_e+h)^2-2t(N)r_e(r_e+h)},
    \end{equation}
    where $t(N)=2\times(\frac{1}{2})^{\frac{1}{N}}-1$.
\end{lemma}
\begin{IEEEproof}
    From Lemma \ref{lem:criticalanalysis} and Theorem \ref{theo:perpro_covpro}, the critical relationship between $N$ and $\gamma$, (\ref{criticalNc}), can be rewritten as $\cos{\gamma^c}=2\times(\frac{1}{2})^{\frac{1}{N}}-1$, that is, $\gamma^c=\arccos{\big(2\times(\frac{1}{2})^{\frac{1}{N}}-1\big)}$. Using the law of cosines, $\cos{\gamma^c}=\frac{r_s^2+r_e^2-d_m^2}{2r_e r_s}$ where $r_s=r_e+h$, we can obtain the critical relationship between the altitude $h$ and the maximum slant range $d_m$ above.
\end{IEEEproof}
The above lemma is an extension of the optimization problem (\ref{design2}), where the optimal value of $\gamma$ is related to $d_m$ and $h$, which are both important parameters of LEO satellite constellations.
\section{Simulation results and discussion}
In this paper, we aim to prove the relationship between coverage probability of users on the sphere and the percolation probability as we defined. The parameters of three existing LEO satellite constellation: Starlink, Oneweb and Kuiper, that we adopt, are shown in Table.\ref{tab:TableOfParam} \cite{osoro2021techno,cakaj2021parameters}.

\begin{figure}
    \centering
    \includegraphics[width=0.8\linewidth]{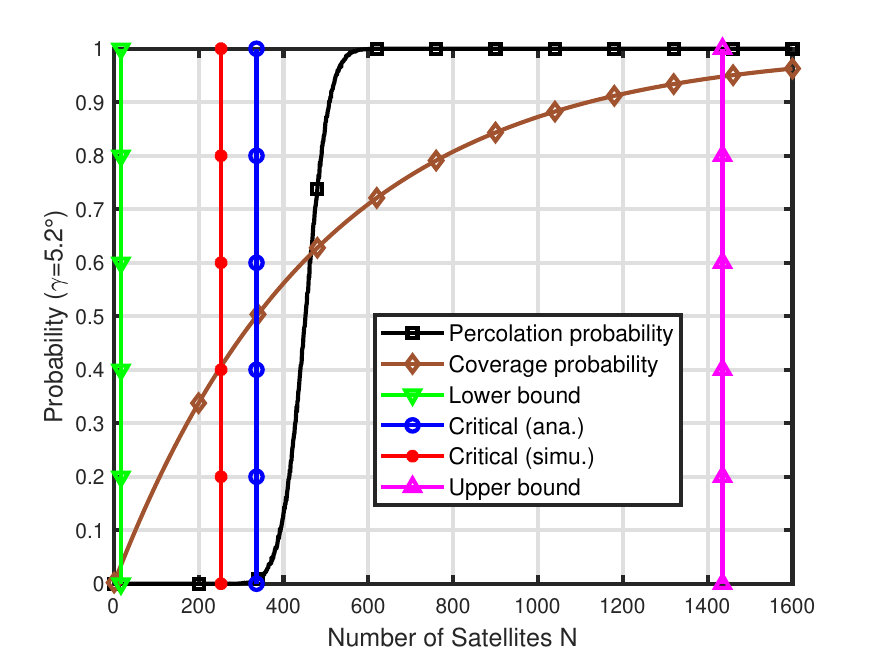}
    \caption{Percolation probability, coverage probability of LEO satellite coverage when $\gamma=5.2$°, with the lower bound $N_L$, upper bound $N_U$, simulated critical value and theoretical value of critical number of LEO satellites $N_c$.}
    \label{fig:gamma52}
\end{figure}

\begin{figure}
    \centering
    \includegraphics[width=0.8\linewidth]{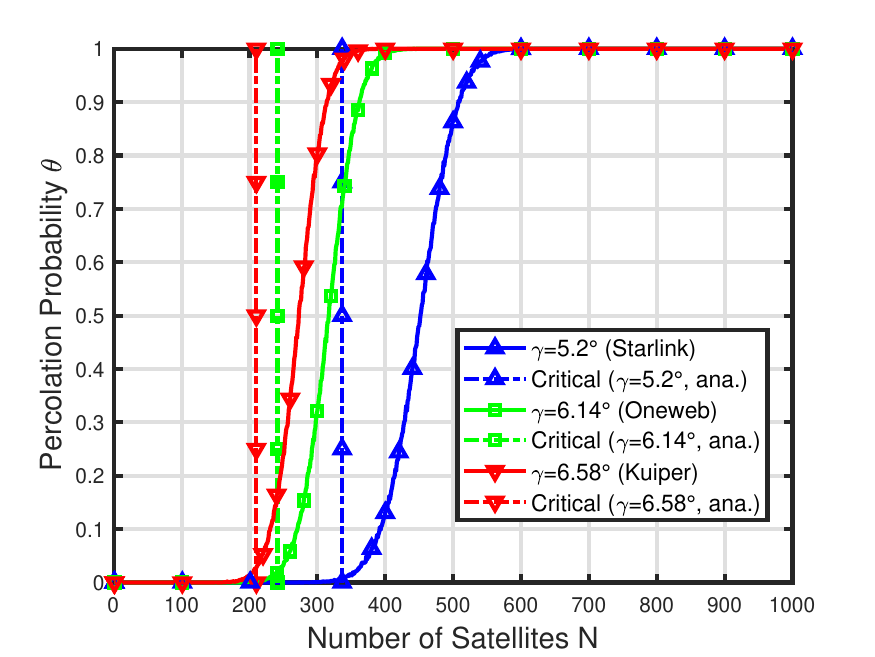}
    \caption{Percolation probability $\theta(N,\gamma)$ versus the number of LEO satellites $N$. Each curve has its corresponding critical number of satellites, $N_c$.}
    \label{fig:companies}
\end{figure}

\begin{figure}
    \centering
    \includegraphics[width=0.8\linewidth]{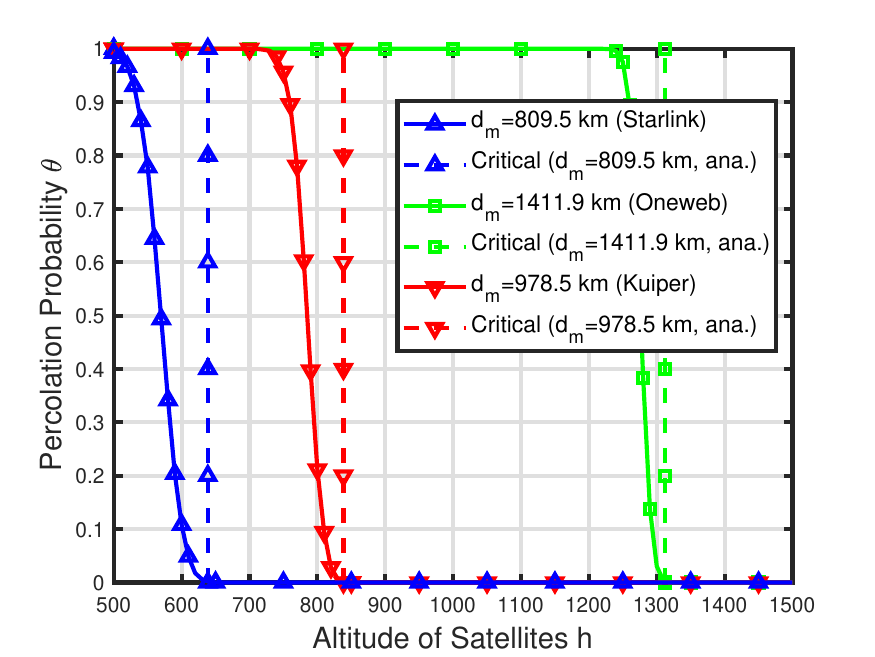}
    \caption{Percolation probability versus the altitude of satellites when $N=500$. Each curve has its corresponding critical constellation altitude $h_c$.}
    \label{fig:Pro_SLR_Altitude}
\end{figure}

\begin{figure}
    \centering
    \includegraphics[width=0.8\linewidth]{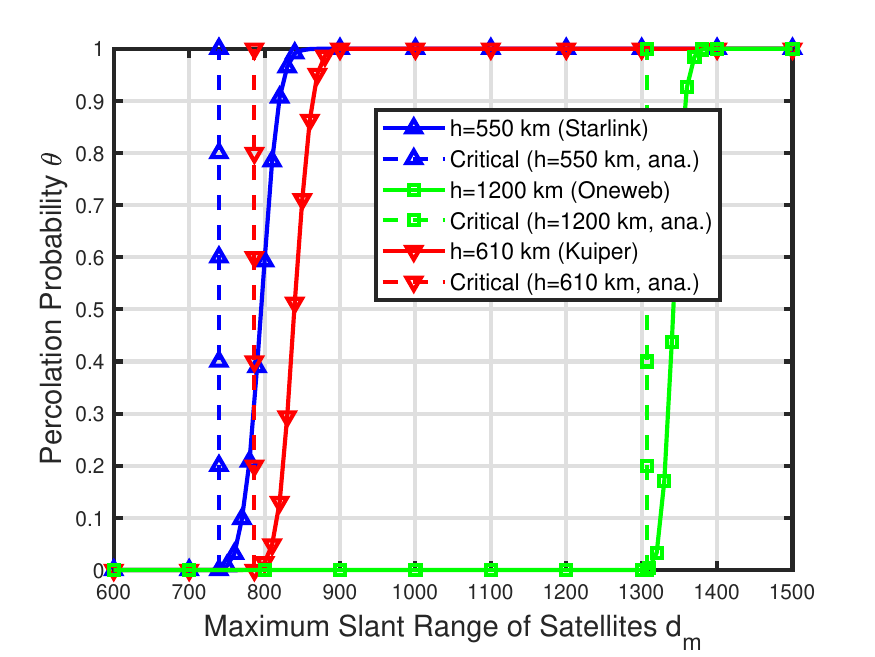}
    \caption{Percolation probability versus the maximum slant range of LEO satellites when $N=500$. Each curve has its corresponding critical maximum slant range $d_m^c$.}
    \label{fig:Pro_Alt_SlantRange}
\end{figure}

\begin{table}[ht]\caption{Parameters of LEO Systems}
\centering
    \begin{tabular}{ {l} | {l} | {l} | {l} }
    \hline
        \hline
    \textbf{Systems} & \textbf{Starlink} & \textbf{Oneweb} & \textbf{Kuiper} \\ \hline
    % \textbf{Number} & 4519 & 648 & 720 \\ \hline
    \textbf{Altitude ($\rm km$)} & 550 & 1200 & 610 \\ \hline
    \textbf{Elevation Angle $\epsilon$ ($^{\circ}$)} & 40.0 & 55.0 & 35.2 \\ \hline
    \textbf{Coverage Angle $\gamma$ ($^{\circ}$)} & 5.20 & 6.14 & 6.58 \\ \hline
    \textbf{Nadir Angle $\eta$ ($^{\circ}$)} & 44.80 & 28.86 & 48.22 \\ \hline
    \textbf{Max Slant Range $d_{m}$ ($\rm km$)} & 809.5 & 1411.9 & 978.5 \\ \hline
    \textbf{Coverage Areas ($\times 10^6\; {\rm km}^2$)} & $1.05$ & $1.46$ & $1.68$ \\ \hline
     \hline
    \end{tabular}
\label{tab:TableOfParam}
%\vspace{-8mm}
\end{table}
 \indent As shown in Fig.\ref{fig:gamma52}, we firstly adopt the Starlink's coverage angle $\gamma=5.2$°. When the number of LEO satellites equals the lower bound (\ref{lowerbound}) in Lemma \ref{lem:lowerbound}, the percolation probability $\theta(N,\gamma)=0$. When the number of LEO satellites equals the upper bound (\ref{upperbound}) in Lemma \ref{lem:upperbound1}, the percolation probability is non-zero. The phase transition of percolation probability is between the lower bound and upper bound. The critical threshold (\ref{criticalNc}) derived in Lemma \ref{lem:criticalanalysis} is slightly higher than the simulated result, but its corresponding percolation probability is extremely low. At this threshold, the coverage probability exceeds 0.5, and percolation probability increases rapidly from a low level (close to 0). This result supports the concept in Theorem \ref{theo:perpro_covpro}.\\
\indent Next, we aim to show the effect of parameters of LEO satellites on the percolation probability. In Fig.\ref{fig:companies}, we adopt the coverage angles $\gamma$ of Starlink, Oneweb and Kuiper. When the number of LEO satellites increases, percolation probability also increases and the derived critical value $N_c$ is the necessary condition for phase transition of percolation probability. For example, Starlink need to provide at least 340 LEO satellites to meet the needs of random large-scale continuous service path, while Oneweb needs 240 LEO satellites and Kuiper needs 200. The critical threshold works well for different values of coverage angle $\gamma$. For realistic applications, the number of LEO satellites also depends on the capacity we need, and our derived critical value is only a necessary condition.\\
\indent Considering 500 LEO satellites, we can observe how the altitude $h$ and maximum slant range $d_m$ of LEO satellites affect the percolation probability together. In Fig.\ref{fig:Pro_SLR_Altitude}, we adopt the maximum slant range $d_m$ of Starlink, Oneweb and Kuiper. When the $h$ increases, the percolation probability decreases because the coverage angle $\gamma$ becomes lower. The critical altitude of LEO satellites for phase transition of percolation probability from non-zero to zero is shown in (\ref{criticalaltitude}). We can notice that these three companies already deploy their LEO satellites at suitable altitudes lower than the critical threshold, where 500 LEO satellites can successfully provide large-scale continuous service for any applications.\\
\indent Similarly, as shown in Fig.\ref{fig:Pro_Alt_SlantRange}, we consider 500 LEO satellites and the altitudes of Starlink, Oneweb and Kuiper constellations. When the $d_m$ increases, percolation probability increases from zero to non-zero due to the increase in nadir $\eta$ and coverage angle $\gamma$. The maximum slant ranges of these three constellations can already support large-scale continuous service. 

In this paper, we verify that the proposed closed-form expression reflects the phase transition behavior of percolation probability when the number of LEO satellite increases. It is worth noting that, the design of constellation is a complex question, where we also need to consider the service strategy and expense. For example, if a considered LEO constellation can only use 100 LEO satellites to provide continuous service for international flights, the required maximum slant range should be higher than the result when $N=500$. We also need to consider the capacity and dynamic selection 
strategy of LEO satellites. In the future, more realistic simulations through orbital propagation tool are expected to be conducted, and multi-layer structure of LEO satellites and the effect from massive users on the traffic should be considered. For example, the kinds of service requirements from IoT devices and mobile users will also lead to different coverage areas and traffic congestion problem of LEO satellite system, which are expected to be solve through routing algorithm and capacity enhancement.
\section{Conclusion}
This paper is the first attempt to show and prove the concept of percolation on the sphere, especially considering the connections between spherical coverage areas. Using the stereographic projection, we defined the percolation on the sphere using the percolation on the projection plane. We first introduced sub-critical and super-critical cases, where the percolation probability $\theta$ is zero and non-zero respectively. We considered two special deployments and derived the lower bound $N_L$ and upper bound $N_U$ of the critical number of LEO satellites $N$. We proved the existence of the critical condition for phase transition of percolation probability from zero to non-zero. Using the hexagonal face percolation on the projection plane, we derived the tight lower bound $N_c^L$ and upper bound $N_c^U$ for percolation, and obtained the closed-form expression of the critical number of satellites $N_c$. We also obtained the expression of critical condition of altitude $h$ and maximum slant range $d_m$. We conducted the simulations to show how these parameters affect the percolation probability and our derived critical expressions can work well to show the phase transition. We emphasized that the critical expressions we derived are the necessary conditions. However, for realistic applications, it is necessary to consider the dynamic selection strategy, capacity and cost.
\appendices
\section{Proof of Lemma~\ref{lem:etagamma}}\label{app:etagamma}
\indent In the triangle $\triangle \yy\oe\textbf{w}_{0,1}$, the maximum slant range $d_m=\|\yy-\textbf{w}_{0,1}\|$. Using the Law of Cosines, we have:
\begin{equation}
    d_m^2+r_s^2-r_e^2=2d_mr_s\cos\eta.
\end{equation}
Therefore, 
\begin{equation}
    d_m=-\sqrt{r_e^2-r_s^2\sin^2\eta}+r_s\cos\eta.
    \label{cosresult}
\end{equation}
Using the Law of Sines, we have:
\begin{equation}
    \frac{d_m}{\sin\gamma}=\frac{r_e}{\sin\eta}.
    \label{sin}
\end{equation}
Substitute (\ref{cosresult}) into (\ref{sin}), we obtain the relationship between $\gamma$ and $\eta$:
   \begin{equation}
       \gamma=\arcsin(\frac{\sin\eta}{r_e}\bigg(-\sqrt{r_e^2-r_s^2\sin^2\eta}+r_s \cos\eta\bigg)).
   \end{equation}

\section{Proof of Theorem \ref{theo:pcov}}\label{app:pcov}
To achieve the coverage probability of each point on the earth, we need to first calculate the area of the spherical cap. As shown in Fig.\ref{fig:caparea},  we can obtain the area of the spherical cap using the integration in polar coordinates:
\begin{equation}
\begin{array}{r@{}l}
    
\mathcal{S}(\gamma)&=\int_{0}^{\gamma}2\pi r(\theta)\cdot r_e\,\dd \theta=\int_{0}^{\gamma}2\pi r_e^2\sin\theta\,\dd \theta\\
&=\displaystyle2\pi r_e^2\cos\theta|_{\gamma}^{0}=\displaystyle2\pi r_e^2(1-\cos\gamma).
\end{array}
\end{equation}

\indent As shown in Fig.\ref{fig:caparea}, the surface area of the spherical cap is equal to the lateral surface area of a cylinder, whose radius is the same as the sphere $r_e$ and height is the same as the spherical cap $H$. This is a classic example of the Mercator projection.\\
\begin{figure}
    \centering
    \includegraphics[width=0.8\linewidth]{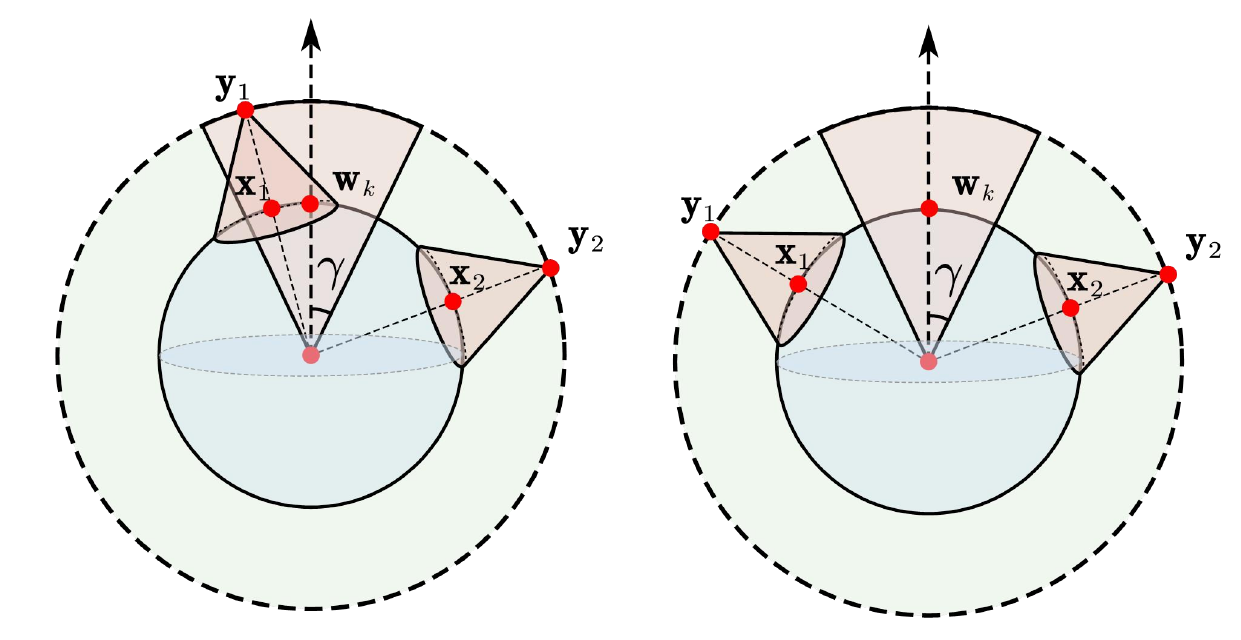}
    \caption{Examples for the event of a point on the sphere located at $\textbf{w}_k$ being covered (left) or not being covered (right) by LEO satellites located at $\yi'$s. }
    \label{fig:covpro}
\end{figure}
\indent As shown on the right of Fig.\ref{fig:covpro}, if the satellite located at $\textbf{y}_1$ is outside the angle range $\gamma$, its coverage area $\mathcal{O}_1=\mathcal{A}_1$ does not include the considered point $\textbf{w}_k$ on the earth. Correspondingly, its coverage center $\textbf{x}_1$ is outside the spherical cap centered at $\textbf{w}_k$, which can be called as the service request area $\mathcal{O}(\textbf{w}_k,\gamma)$. Therefore, the probability of the point located at $\textbf{w}_k$ not being covered by the LEO satellites located at $\yi'$s is:
\begin{equation}%\displaystyle
\begin{array}{r@{}l}
    p&_{\rm ncov}(N,\gamma)\triangleq \mathbb{P}\{\textbf{w}_k {\rm\,is\,not\,covered\,by\,}\yi'{\rm s}\}\\
    &\overset{(a)}{=} \prod_{i=1}^{N}\mathbb{P}\{\textbf{w}_k {\rm\,is\,not\,covered\,by\,}\yi\}\\
    &\overset{(b)}{=}\big(1-\frac{\mathcal{S(\gamma)}}{4\pi r_e^2}\big)^{N}=\big(1-\frac{2\pi r_e^2 (1-\cos\gamma)}{4\pi r_e^2}\big)^{N}\\
    &=\big(\frac{1+\cos\gamma}{2}\big)^{N},\\
\end{array}
\end{equation}
% \\
%     &
where steps (a) and (b) both come from the BPP assumption of LEO satellite deployment. As shown on the left of Fig.\ref{fig:covpro}, if there is at least one LEO satellite located inside $\mathcal{O}(\textbf{w}_k,\gamma)$ , the considered point $\textbf{w}_k$ on the earth is covered. Therefore, the coverage probability of each point on the earth can be defined as:
\begin{equation}
\begin{array}{r@{}l}
    p&_{\rm cov}(N,\gamma)\triangleq \mathbb{P}\{\textbf{w}_k {\rm\,is\,covered\,by\,at\,least\,one\,of\,}\yi'{\rm s}\}\\
    &=1-\mathbb{P}\{\textbf{w}_k {\rm\,is\,not\,covered\,by\,}\yi'{\rm s}\}\\
    &=1-\big(\frac{1+\cos\gamma}{2}\big)^{N}.
\end{array}
\end{equation}

\section{Proof of Lemma~\ref{lem:circularradius}}\label{app:circularradius}
\begin{figure}
    \centering
    \includegraphics[width=0.75\linewidth]{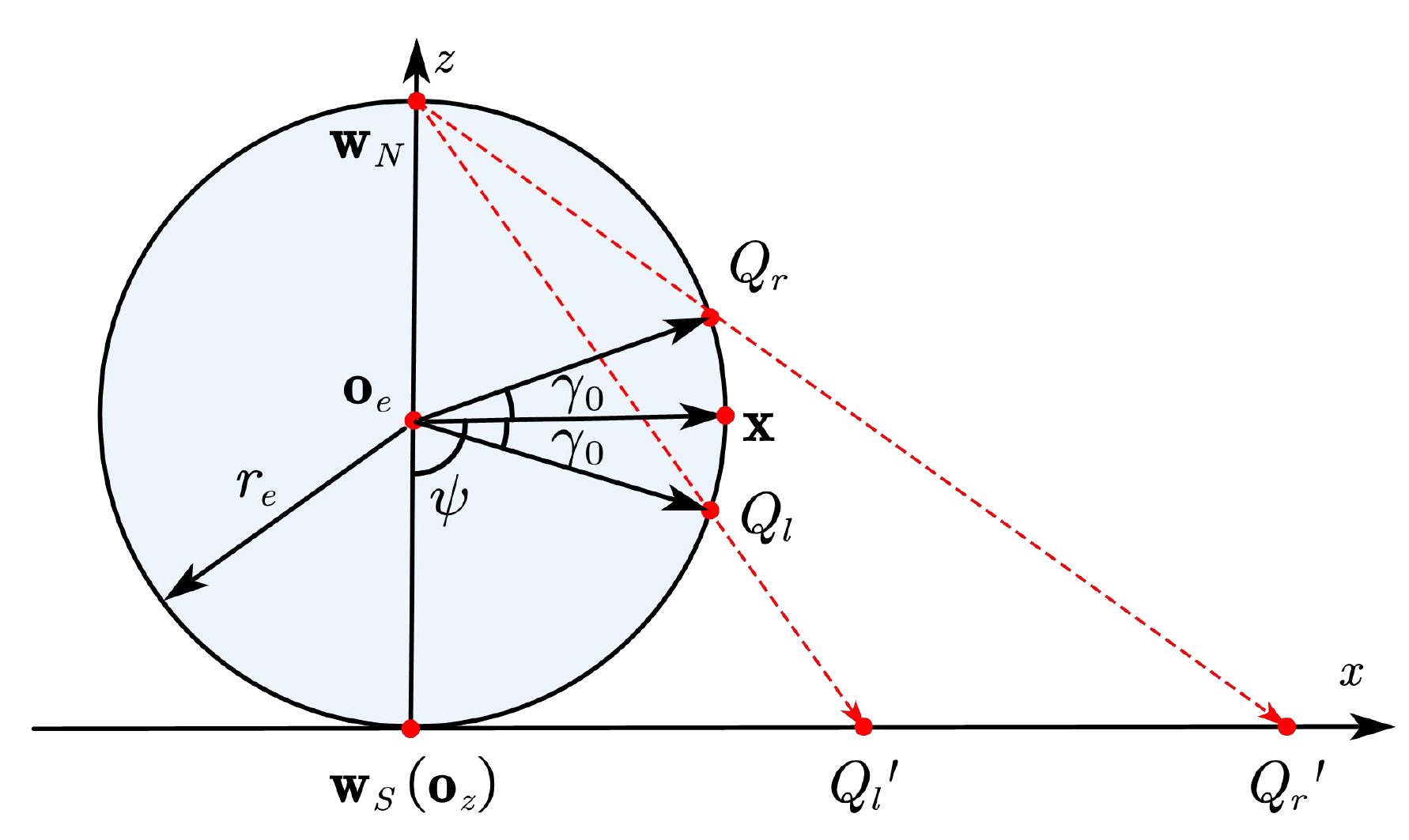}
    \caption{The stereographic projection $Q'_{l}=\mathcal{F}(Q_{l})$ and $Q'_{r}=\mathcal{F}(Q_{r})$. }
    \label{fig:OQOQ}
\end{figure}

\indent Assume that $\psi=\angle\textbf{w}_S \oe \textbf{x}$ and $\gamma_0$ is the central angle of a spherical cap. The diameter of the projection and its corresponding arcs are both located at the plane $xoz$. First consider the case where $0\leq \psi< \pi-\gamma_0$. As shown in Fig.\ref{fig:OQOQ}, the points at both ends of the diameters satisfy the projection relationship
:
\begin{equation}
    Q'_{l}=\mathcal{F}(Q_{l}),\,Q'_{r}=\mathcal{F}(Q_{r}).
\end{equation}
Therefore, the radius of projected circular area satisfies:
\begin{equation}
    r(\psi)=\frac{1}{2}\bigg|\|\textbf{o}_z Q'_{r}\|-\|\textbf{o}_z Q'_{l}\|\bigg|.
\label{RQQ}
\end{equation}
\indent We have
\begin{equation}
\begin{array}{r@{}l}
    \|\textbf{o}_z& Q'_{r}\|\displaystyle=2r_e \tan(\angle \textbf{w}_S \textbf{w}_N Q'_{r})\\
    &\displaystyle=2r_e \tan(\frac{\angle \textbf{w}_S \textbf{o}_e Q_{r}}{2})\displaystyle=2r_e \tan(\frac{\psi+\gamma_0}{2}).
\end{array}
\label{0Qr}
\end{equation}
and 
\begin{equation}
\begin{array}{r@{}l}
    \|\textbf{o}_z& Q'_{l}\|\displaystyle=2r_e \tan(\angle \textbf{w}_S \textbf{w}_N Q'_{l})\\
    &\displaystyle=2r_e \tan(\frac{\angle \textbf{w}_S \textbf{o}_e Q_{l}}{2})\displaystyle=2r_e \tan(\frac{\psi-\gamma_0}{2}).
\end{array}
\label{0Ql}
\end{equation}
\indent Substitute (\ref{0Qr}) and (\ref{0Ql}) into (\ref{RQQ}), we obtain:
\begin{equation}
    r(\psi)=r_e\bigg|\tan(\frac{\psi+\gamma_0}{2})-\tan(\frac{\psi-\gamma_0}{2})\bigg|.    
\label{rpsi}
\end{equation}
% \indent It is worth noting that this equation is also suitable for the case where $\pi-\gamma<\phi_i\leq\pi$. When $\psi_i=\pi-\gamma$, the radius goes to $+\infty$, therefore, $r_{i}(\psi_i)<+\infty$. Next, we prove that there exists the minimum value of $r_{i}(\psi_i)$.\\ 
% \indent First consider the case where $0\leq\psi_i <\pi-\gamma$. Taking the derivative of $r_{i}(\psi_i)$ with respect to $\psi_i$, we have
% \begin{equation}
% \begin{array}{r@{}l}
%     &\displaystyle\frac{\partial r_{i}(\psi_i)}{\partial \psi_i}=\displaystyle r_e\bigg(\frac{1}{2\cos^2(\frac{\psi_i+\gamma}{2})}-\frac{1}{2\cos^2(\frac{\psi_i-\gamma}{2})}\bigg)\\
%     &=\displaystyle r_e\bigg(\frac{1}{1+\cos(\psi_i+\gamma)}-\frac{1}{1+\cos(\psi_i-\gamma)}\bigg)\geq 0.
% \end{array}
% \end{equation}
% Similarly, for the case where $\pi-\gamma<\psi_i\leq\pi$, 
% \begin{equation}
% \begin{array}{r@{}l}
%     &\displaystyle\frac{\partial r_{i}(\psi_i)}{\partial \psi_i}=\displaystyle r_e\bigg(\frac{1}{2\cos^2(\frac{\psi_i-\gamma}{2})}-\frac{1}{2\cos^2(\frac{\psi_i+\gamma}{2})}\bigg)\\
%     &=\displaystyle r_e\bigg(\frac{1}{1+\cos(\psi_i-\gamma)}-\frac{1}{1+\cos(\psi_i+\gamma)}\bigg)\leq 0.
% \end{array}
% \end{equation}
% Therefore, the minimum value of $r_{i}(\psi_i)$ is:
% \begin{equation}
% \begin{array}{r@{}l}
%     R_{z,min}&=\min\{r_{i}(0),r_{i}(\pi)\}\\
%     &=\min\{2r_e \tan (\frac{\gamma}{2}),2r_e \cot (\frac{\gamma}{2})\}
% \end{array}
% \end{equation}
%  Therefore, $R_{z,\min}=2r_e \tan(\frac{\gamma}{2})$ and the range of $r_{i}(\psi_i)$ is:
When the central angle $\gamma_0$ is less than $\frac{\pi}{2}$, (\ref{rpsi}) works for $\gamma_0-\pi<\psi<\pi-\gamma_0$. The range of the radius of the projected circular area is:
\begin{equation}
    2r_e \tan(\frac{\gamma_0}{2})\leq r(\psi)<+\infty.
\end{equation}
Conversely, if the radius of the projected circular area is $r$, the range of the central angle of the original spherical cap, $\gamma_0$, is:
\begin{equation}
    0<\gamma_0\leq 2\arctan(\frac{r}{2r_e}).
\label{gamma0range}
\end{equation}

% \section{Proof of Theorem~\ref{theo:concomsphere}}\label{app:concomsphere}
% In this appendix, we need to prove the concept of percolation on the sphere.

% \section{Proof of Lemma~\ref{lem:lowerbound}} \label{app:lowerbound}
% Here, we need to prove the lower bound.
% \section{Proof of Lemma~\ref{lem:upperbound1}} \label{app:upperbound1}
% Here, we need to prove the upper bound.
\section{Proof of Lemma~\ref{lem:nondecreasing}\label{app:nondecreasing}}
To prove that there exists a critical number of LEO satellites that causes the phase transition of percolation probability, we need to first prove that the percolation probability is a non-decreasing function of $N$ when the $\gamma$ is fixed. \\
\indent Firstly, we assume that the LEO satellites are deployed at the same altitude with the same nadir angle $\eta$, therefore, the coverage angle $\gamma$ is also fixed. The locations of satellites follow a BPP around the earth at a certain altitude. Consider two sets of satellites $\Phi_1$ and $\Phi_2$ with the number of vertices $N_1$ and $N_2$, respectively, where $N_1<N_2$. Since $\Phi_1$ and $\Phi_2$ are both BPPs, $\Phi_1$ can be constructed by removing any one vertice inside $\Phi_2$. Similarly, $\Phi_2$ can be constructed by adding any other vertice into $\Phi_1$. Because we discuss their coverage areas on the sphere, the set of spherical caps' centers $V_1$ and $V_2$ can be generated following BPPs with number $N_1$ and $N_2$ at the altitude $0\;\rm km$ (on the sphere), where $V_1\subseteq V_2$ when $\Phi_1\subseteq\Phi_2$.\\
\indent Removing vertice from $\Phi_2$ or adding vertice into $\Phi_1$ both lead to the change of the edge set, where $E_1\subseteq E_2$. The connected components in these two random graphs are defined as: $K_{x,1}\subseteq G_{x}(V_{x,1},V_{x,1})$ and $K_{x,2}\subseteq G_{x}(V_{x,2},V_{x,2})$, which satisfy $K_{x,1}\subseteq K_{x,2}$. Therefore, $0<N_1<N_2$ indicates $\theta(N_1,\gamma)\leq \theta(N_2,\gamma)$, \ie the percolation probability is a non-decreasing function of $N$.\\
\indent It is worth noting that, in this paper, we use two kinds of projection: \textit{i) mapping the satellite locations to the sphere}, which obtains the coverage centers, \textit{ii) mapping every point on the sphere to the projection plane}, which helps us define the percolation on the sphere. The mapping from satellites to coverage centers keep the BPP properties, and mapping the sphere to the projection plane keeps almost all the connections between coverage areas, which has been introduced in the Lemma \ref{lem:mappingrelation}. Therefore, the projections of connected components on the considered projection plane also satisfy $K_{z,1}\subseteq K_{z,2}$. On the plane, we can focus on the connected component containing the origin $\textbf{o}_z$, \ie $K_z(0)$. Therefore, the percolation probabilities for these two cases also satisfy $\P(|K_{z,1}(0)|=\infty)\leq \P(|K_{z,2}(0)|=\infty)$.
\section{Proof of Lemma \ref{lem:upperbound1}}
\label{app:upperbound}
Because the area of a sphere is finite, the graph percolates once the whole sphere is covered by LEO satellites. Therefore, the percolation probability can be lower bounded by the full coverage probability, \ie
\begin{equation}
    \theta(N,\gamma)\geq \P\{{\rm Full\;Coverage}|N,\gamma\}.
\label{full}
\end{equation}
\indent We have proposed our `full coverage scheme' in Sec.\ref{subsec:phasetransition}. Because it is one way to realize full coverage, the probability of a successful deployment is less than or equal to the full coverage probability, \ie
\begin{equation}
\begin{array}{r@{}l}
    \P&\{{\rm Full\;Coverage}|N,\gamma\}\\
    &\geq \P\{{\rm The\; Proposed\;Full\;Coverage\;Scheme}|N,\gamma\}.
\end{array}
\label{successfull}
\end{equation}
Therefore, if we can find a computable and non-zero value of the probability of our proposed full coverage scheme, the percolation probability is proved to be non-zero. Firstly, we need to prove that full coverage can be realized under such a scheme. \\
\indent In Fig.\ref{fig:Upperbound}, we propose to represent the whole sphere using the union of sphere caps and introduce the steps we need. Let $\Omega$ denote the whole sphere and $\mathcal{A}_i$ denote the coverage area of the \textit{i}th LEO satellite. Therefore, the full coverage defined as an event: $\Omega\subseteq \bigcup_{i=1}^{N}\mathcal{A}_i$.\\
\indent We first uniformly divide the whole sphere into $2m$ `slices' using $2m$ meridians, where the \textit{j}th slice is denoted by $\mathcal{S}_j$ and spans $\frac{\pi}{m}$ of longitude. The whole sphere can be expressed as the union of $\mathcal{S}_j$'s, that is, $\Omega=\bigcup_{j=1}^{2m}\mathcal{S}_j$.\\
\indent On the sphere, we assume that slices $\mathcal{S}_j$ and $\mathcal{S}_{m+j}$ are symmetric about the earth's center. They can be contained by a `belt' defined as $\mathcal{T}_j$, \ie $\mathcal{S}_j\bigcup\mathcal{S}_{m+j}\subseteq \mathcal{T}_j$. Therefore, the union of slices is the subset of the union of belts, \ie $\bigcup_{j=1}^{2m}\mathcal{S}_j\subseteq \bigcup_{j=1}^{m}\mathcal{T}_j$. Because $\Omega=\bigcup_{j=1}^{2m}\mathcal{S}_j$ and $\bigcup_{j=1}^{m}\mathcal{T}_j\subseteq \Omega$, the union of the belts is the same as the whole sphere, \ie $\Omega=\bigcup_{j=1}^{m}\mathcal{T}_j$.\\
\indent Next, we can rotate the belt $\mathcal{T}_1$ and make it symmetric about the equatorial plane. Belt $\mathcal{T}_1$ spans $\frac{\pi}{m}$ of latitude. It is able to uniformly divide $\mathcal{T}_1$ into $n$ pieces using $n$ meridians, \ie $\mathcal{T}_1=\bigcup_{k=1}^{n}\mathcal{D}_{1,k}$. where each piece $\mathcal{D}_{1,k}$ spans $\frac{\pi}{m}$ of latitude and $\frac{2\pi}{n}$ of longitude. Other belts can be divided in the same way, and all pieces have the same shape and size. Therefore, the whole sphere is the same as the union of such `pieces', \ie $\Omega=\bigcup_{j=1}^{m}\bigcup_{k=1}^{n}\mathcal{D}_{j,k}$.\\
\indent We consider the piece $\mathcal{D}_{1,1}$ which spans $\frac{\pi}{m}$ of latitude and $\frac{2\pi}{n}$ of longitude firstly. We can find the minimum spherical cap $\mathcal{E}_{1,1}$ containing $\mathcal{D}_{1,1}$, \ie $\mathcal{D}_{1,1}\subseteq \mathcal{E}_{1,1}$. The spherical cap has the same center of $\mathcal{D}_{1,1}$, and its central angle $\zeta$ satisfies:
\begin{equation}
    \zeta =\arccos{\bigg(\cos{\frac{\pi}{2m}}\cos{\frac{\pi}{n}}\bigg)}.
\end{equation}
For each piece $\mathcal{D}_{j,k}$, we can find its corresponding spherical cap $\mathcal{E}_{j,k}$ with the same central angle $\zeta$, where $\mathcal{D}_{j,k}\subseteq \mathcal{E}_{j,k}$. Therefore, the union of pieces is the subset of the union of these spherical caps, \ie $\bigcup_{j=1}^{m}\bigcup_{k=1}^{n}\mathcal{D}_{j,k}\subseteq \bigcup_{j=1}^{m}\bigcup_{k=1}^{n}\mathcal{E}_{j,k}$. Because $\bigcup_{j=1}^{m}\bigcup_{k=1}^{n}\mathcal{E}_{j,k}\subseteq \Omega$ and $\Omega=\bigcup_{j=1}^{m}\bigcup_{k=1}^{n}\mathcal{D}_{j,k}$, we obtain: $\Omega=\bigcup_{j=1}^{m}\bigcup_{k=1}^{n}\mathcal{E}_{j,k}$.\\
\indent To realize the full coverage, we aim to ensure that each spherical cap $\mathcal{E}_{j,k}$ is covered. We can follow three steps: i) make $m$ and $n$ large enough to make the spherical cap $\mathcal{E}_{j,k}$ can be covered by one satellite whose coverage angle is $\gamma$, \ie $\zeta< \gamma$, ii) use $m\times n$ LEO satellites and deploy them one by one, \ie $\mathcal{E}_{j,k}\subseteq \mathcal{A}_{n(j-1)+k}$. In this case, we can ensure that $\Omega=\bigcup_{j=1}^{m}\bigcup_{k=1}^{n}\mathcal{A}_{n(j-1)+k}$ because $\bigcup_{j=1}^{m}\bigcup_{k=1}^{n}\mathcal{A}_{n(j-1)+k}\subseteq \Omega$ and $\bigcup_{j=1}^{m}\bigcup_{k=1}^{n}\mathcal{E}_{j,k}\subseteq \bigcup_{j=1}^{m}\bigcup_{k=1}^{n}\mathcal{A}_{n(j-1)+k}$.\\
\indent To make $\zeta<\gamma$, we can design $m$ first and then $n$. The requirements are: i) $\frac{\pi}{2m}<\gamma$ and $\frac{\pi}{n}<\gamma$ and ii) $\zeta = \arccos{(\cos{\frac{\pi}{2m}}\cos{\frac{\pi}{n}})}<\gamma$, that is, $\cos{\frac{\pi}{2m}}\cos{\frac{\pi}{n}}>\cos{\gamma}$. We can first choose a feasible $m$, for example, $m=\left \lceil \frac{\pi}{\gamma} \right \rceil$ to let the inequality $\frac{\pi}{2m}<\gamma$ hold. Next, because $\cos{\frac{\pi}{n}}>\max\{\cos{\gamma},\frac{\cos{\gamma}}{\cos{\frac{\pi}{2m}}}\}$,  $\frac{\pi}{n}<\arccos{\frac{\cos{\gamma}}{\cos{\frac{\pi}{2m}}}}$, the choice of $n$ should satisfy $n>\left \lceil \frac{\pi}{\arccos{\frac{\cos{\gamma}}{\cos{\frac{\pi}{2m}}}}} \right \rceil$. Therefore, we can let $n=\left \lceil \frac{\pi}{\arccos{\frac{\cos{\gamma}}{\cos{\frac{\pi}{2m}}}}} \right \rceil+1$ and the total number of satellites is $N_U=m\times n$. Therefore, when $N=N_U$, we can realize the full coverage on the sphere.\\
\indent Based on such a special deployment, we can derive the lower bound of its probability. We assume that the center of the first satellite's coverage area $\mathcal{A}_{1,1}$ is $\textbf{x}_{1,1}$ and the center of the spherical cap $\mathcal{E}_{1,1}$ is $\textbf{w}_{1,1}$. When $\angle \textbf{x}_{1,1} \textbf{o}_e \textbf{w}_{1,1}<\gamma-\zeta$, $\mathcal{E}_{1,1}$ is totally covered by $\mathcal{A}_{1,1}$. The probability of such an event is $\frac{1-\cos{(\gamma-\zeta)}}{2}$. We can deploy other satellites in the same way to ensure all $\mathcal{E}_{j,k}$'s are covered. The probability of such a successful full deployment is:
\begin{equation}
\begin{array}{r@{}l}
    \P&\{{\rm The\; Proposed\;Full\;Coverage\;Scheme}|N=N_U,\gamma\}\\
    &=\big(\frac{1-\cos{(\gamma-\zeta)}}{2}\big)^N,
\end{array}
\end{equation}
which is computable and non-zero. Therefore, using the inequalities (\ref{full}) and (\ref{successfull}), we prove that the percolation probability is strictly larger than 0 when $N=N_U$. Similarly, when $N>N_U$, we can deploy the first $N_U$ satellites in the same way, and deploy the other satellites randomly.
Therefore, for $N\geq N_U$, percolation probability is always non-zero, \ie
\begin{equation}
    \theta(N,\gamma)>0\;{\rm for}\;N\geq N_U.
\end{equation}

\section{Proof of Lemma \ref{lem:phasetransition}}\label{app:phasetransition}
\indent According to Lemma \ref{lem:lowerbound} and \ref{lem:upperbound1}, we know that: i) $\theta(N,\gamma)=0$ for $N\leq N_L$ and ii) $\theta(N,\gamma)>0$ for $N\geq N_U$. Because the percolation probability $\theta(N,\gamma)$ is a non-decreasing function of $N$, there must exist a critical value of $N$, \ie $N_c$, which satisfies:
\begin{equation}
    \begin{array}{c}
        \theta(N,\gamma)=0,\, {\rm for}\; N< N_c,\\
        \theta(N,\gamma)>0,\, {\rm for}\; N> N_c. 
    \end{array}
\end{equation}
% It is worth noting that $N$ should be an integer but $N_c$ does not need to be defined as an integer. Therefore, we can define the $N_c$ using:
% \begin{equation}
%     \begin{array}{c}
%         \theta(N,\gamma)=0,\, {\rm for}\; N\leq \left\lfloor N_c\right\rfloor,\\
%         \theta(N,\gamma)>0,\, {\rm for}\; N\geq \left\lceil N_c\right\rceil. 
%     \end{array}
% \end{equation}
The upper and lower bounds should satisfy $N_L\leq \left\lfloor N_c\right\rfloor$ and $\left\lceil N_c\right\rceil\leq N_U$. However, these two inequalities do not hold equality at the same time because the percolation probability can not be zero and non-zero at the same time.
\section{Proof of Theorem \ref{theo:inhomohexagon}}\label{app:inhomohexagon}
\indent Assume that different hexagons have different probabilities of being open or closed and both of them have their minimum value $p_{\rm cov}^{\min}$ and $p_{\rm ncov}^{\min}$, \ie $\P\{\mathcal{H}_{l} {\rm \; is\;open}\} \geq p_{\rm cov}^{\min}$ and $\P\{\mathcal{H}_{l} {\rm \; is\;closed}\} \geq p_{\rm ncov}^{\min}$.\\ %, where
% \begin{equation}
%     \P\{\mathcal{H}_{l} {\rm \; is\;open}\} \geq p_{\rm cov}^{\min}
% \end{equation}
% and
% \begin{equation}
%     \P\{\mathcal{H}_{l} {\rm \; is\;closed}\} \geq p_{\rm ncov}^{\min}.
% \end{equation}
\indent We firstly discuss the case where the $\P\{\mathcal{H}_{l} {\rm \; is\;open}\}$ has its minimum value $p_{\rm cov}^{\min}$. Considering the hexagons with side length $a$, we can cover the hexagons following their coverage probability $\P\{\mathcal{H}_{l} {\rm \; is\;open}\}$. The random graph in this case is defined as $G_{z}^{\rm cov}$. The probabilities $p_{\rm cov}^{\min}$ and $\P\{\mathcal{H}_{l} {\rm \; is\;open}\}$ are assumed larger than 0. \\
\indent Define the random graph of `open hexagonal faces' generated by probability $p_{\rm cov}^{\min}$ as $G_{z,\min}^{\rm cov}$. We can generate the $G_{z,\min}^{\rm cov}$ through removing each open face $\mathcal{H}_{l}$ in $G_{z}^{\rm cov}$ by probability $p_{1}^l=1-p_{\rm cov}^{\min}/\P\{\mathcal{H}_{l} {\rm \; is\;open}\}$ where $p_{1}^l\in [0,1)$. Therefore, all open faces $G_{z,\min}^{\rm cov}$ are contained by $G_{z}^{\rm cov}$, \ie $G_{z,\min}^{\rm cov}\subseteq G_{z}^{\rm cov}$. When $\P\{\mathcal{H}_{l} {\rm \; is\;open}\}>1/2$, $p_{\rm cov}^{\min}>1/2$, the percolation probability $\P\{|G_{z,\min}^{\rm cov}|=\infty\}>0$, and the percolation probability of $G_z^{\rm cov}$ also satisfies $\P\{|G_{z}^{\rm cov}|=\infty\}>0$. In conclusion, the sufficient condition for non-zero percolation probability is 
\begin{equation}
    \P\{\mathcal{H}_{l} {\rm \; is\;open}\}>1/2.
\end{equation}
\indent Similarly, we define the random graph of `closed hexagonal faces' generated by probability $p_{\rm ncov}^{\min}$ as $G_{z,\min}^{\rm ncov}$. We can generate the $G_{z,\min}^{\rm ncov}$ through removing each closed face $\mathcal{H}_{l}$ in $G_{z}^{\rm ncov}$ by probability $p_{2}^l=1-p_{\rm ncov}^{\min}/\P\{\mathcal{H}_{l} {\rm \; is\;closed}\}$ where $p_{2}^l\in [0,1)$. Therefore, all closed faces $G_{z,\min}^{\rm ncov}$ are contained by $G_{z}^{\rm ncov}$, \ie $G_{z,\min}^{\rm ncov}\subseteq G_{z}^{\rm ncov}$. When $\P\{\mathcal{H}_{l} {\rm \; is\;closed}\}>1/2$, $p_{\rm ncov}^{\min}>1/2$, the percolation probability $\P\{|G_{z,\min}^{\rm ncov}|=\infty\}>0$, and the percolation probability of $G_z^{\rm ncov}$ also satisfies $\P\{|G_{z}^{\rm ncov}|=\infty\}>0$. In conclusion, the sufficient condition for zero percolation probability is 
\begin{equation}
    \P\{\mathcal{H}_{l} {\rm \; is\;closed}\}>1/2.
\end{equation}
\section{Proof of Lemma \ref{lem:boundsforhexagons}}\label{app:boundsforhexagons}
\indent We first consider the hexagon $\mathcal{H}_{l}$ with the side length 
$a$. The circular area $\tilde{\mathcal{O}}_{l}$ with radius $a$ has the same center as $\mathcal{H}_{l}$. The center of 
$\mathcal{F}^{-1}(\tilde{\mathcal{O}}_{l})$ is $\textbf{x}_{o,l}$ and the center of $\mathcal{A}_i$ is $\textbf{x}_i$. The probability each hexagonal face $\mathcal{H}_{l}$ being closed satisfy:
\begin{equation}%\displaystyle
\begin{array}{r@{}l}
    \P&\displaystyle\{\mathcal{H}_{l}\;{\rm is\;closed}\}\\
    &=\P\{\mathcal{H}_{l}\;{\rm is\;not\;covered\;by\;}\bigcup_{i=1}^{N}\mathcal{F}(\mathcal{A}_i)\}\\
    &\geq\P\{\tilde{\mathcal{O}}_{l}\;{\rm is\;not\;covered\;by\;}\bigcup_{i=1}^{N}\mathcal{F}(\mathcal{A}_i)\}\\
    % &=\P\displaystyle\{\mathcal{F}^{-1}(\tilde{\mathcal{O}}_{l})\;{\rm is\;not\;covered\;by\;}\bigcup_{i=1}^{N}\mathcal{A}_i\}\\
    &\geq\prod_{i=1}^{N}\P\{\mathcal{F}^{-1}(\tilde{\mathcal{O}}_{l})\;{\rm is\;not\;covered\;by\;}\mathcal{A}_i\}\\
    &\geq\prod_{i=1}^{N}\P\{\angle \textbf{x}_{o,l} \textbf{o}_e \textbf{x}_i>\gamma+\gamma_m\}\\
    &=\big(\frac{1+\cos(\gamma+\gamma_m)}{2}\big)^{N},\\
\end{array}
\end{equation}
where $\gamma_m$ is the maximum central angle of the original spherical cap of hexagon's minimum circumscribed circle. Because the radius of the minimum circumscribed circle is $a$, from (\ref{gamma0range}), we can obtain its expression:
\begin{equation}
    \gamma_m=2\arctan \frac{a}{2r_e}.
\end{equation}
% \indent Next, we consider another circular area $\tilde{\mathcal{O}}_{l}^{\rm inner}$ with radius $\frac{\sqrt{3}}{2}a$ has the same center as $\mathcal{H}_{l}$. The center of $\mathcal{F}^{-1}(\tilde{\mathcal{O}}_{l}^{\rm inner})$ is also $\textbf{x}_{o,l}$ and the center of $\mathcal{A}_i$ is $\textbf{x}_i$. 
Similarly, the probability of $\mathcal{H}_{l}$ being covered satisfy:
\begin{equation}
\begin{array}{r@{}l}
    \P&\displaystyle\{\mathcal{H}_{l}\;{\rm is\;open}\}\\
    &=\P\{\mathcal{H}_{l}\;{\rm is\;covered\;by\;}\bigcup_{i=1}^{N}\mathcal{F}(\mathcal{A}_i)\}\\
    &\geq\P\{\tilde{\mathcal{O}}_{l}\;{\rm is\;covered\;by\;}\bigcup_{i=1}^{N}\mathcal{F}(\mathcal{A}_i)\}\\
    % &=\P\displaystyle\{\mathcal{F}^{-1}(\tilde{\mathcal{O}}_{l})\;{\rm is\;covered\;by\;}\bigcup_{i=1}^{N}\mathcal{A}_i\}\\
    &\geq\P\{\mathcal{F}^{-1}(\tilde{\mathcal{O}}_{l})\;{\rm is\;covered\;by\;at\; least\;one\;of\;}\mathcal{A}_i\}\\
    & =1-\prod_{i=1}^{N}\P\{\mathcal{F}^{-1}(\tilde{\mathcal{O}}_{l})\;{\rm is\;not\;covered\;by\;}\mathcal{A}_i\}\\
    &\geq1-\prod_{i=1}^{N}\P\{\angle \textbf{x}_{o,l} \textbf{o}_e \textbf{x}_i>\gamma-\gamma_m\}\\
    &=1-\big(\frac{1+\cos(\gamma-\gamma_m)}{2}\big)^{N}.\\
\end{array}
\end{equation}
% where
% \begin{equation}
%     \gamma_m=2\arctan \frac{\sqrt{3}a}{2 r_e}.
% \end{equation}
% We assume that the side length $a$ is much smaller than the coverage radius of satellites on the plane, $\gamma_m$ and $\gamma_m$ are both assumed much less than $\gamma$, which is the coverage angle of each LEO satellite.
\indent We assume that the side length $a$ is much smaller than the coverage radius of satellites on the plane, $\gamma_m$ is assumed much less than the coverage angle $\gamma$ of each LEO satellite.
\section{Proof of Lemma \ref{lem:criticalanalysis}}\label{app:criticalanalysis}
\indent Notice that, the upper bound \begin{equation}
\displaystyle N_c^U=\displaystyle\frac{\ln 2}{\ln 2-\ln(1+\cos(\gamma-2\arctan \frac{a}{2r_e}))}
\end{equation}
can be considered as an increasing function of $a$ and the lower bound
\begin{equation}
    N_c^L=\frac{\ln 2}{\ln 2-\ln(1+\cos(\gamma+2\arctan \frac{a}{2r_e}))}
\end{equation}
can be considered as a decreasing function of $a$. When the side length $a$ approaches 0, the limit values of the upper bound $N_c^{U}$ and lower bound $N_c^{L}$ are both approach to the same value:
\begin{equation}
\begin{array}{r@{}l}
\lim\limits_{a\rightarrow 0^+} N_c^U&=\lim\limits_{a\rightarrow 0^+}\frac{\ln 2}{\ln 2-\ln(1+\cos(\gamma-2\arctan \frac{a}{2r_e}))}\\
&=\frac{\ln 2}{\ln 2-\ln(1+\cos\gamma)}
\end{array}
\end{equation}
and 
\begin{equation}
\begin{array}{r@{}l}
\lim\limits_{a\rightarrow 0^+} N_c^L&=\lim\limits_{a\rightarrow 0^+}\frac{\ln 2}{\ln 2-\ln(1+\cos(\gamma+2\arctan \frac{a}{2r_e}))}\\
&=\frac{\ln 2}{\ln 2-\ln(1+\cos\gamma)}
\end{array}
\end{equation}

Because $N_c^{L}\leq N_c\leq N_c^U$, the limit value of $N_c$ should be the same as them, \ie
\begin{equation}
    N_c=\displaystyle\frac{\ln 2}{\ln 2-\ln(1+\cos\gamma)}.
\end{equation}

This is also the closed-form expression of critical number of LEO satellites which is always located between these two bounds. %Because the limit value of $N_c^L$ and $N_c^U$ are the same, the upper and lower bounds obtained through this stereographic projection method are both tight. However, they can not be used directly because we focus on continuous percolation on the sphere rather than the hexagonal lattic design on the plane, therefore, we only choose $a=0$ and adopt the critical number $N_c$.\\

\ifCLASSOPTIONcaptionsoff
  \newpage
\fi

\bibliographystyle{IEEEtran}
% Generated by IEEEtran.bst, version: 1.14 (2015/08/26)

% \bibliography{ref}

% \begin{IEEEbiography}
% % [{\includegraphics[width=1in,height=1.25in,clip,keepaspectratio]{./AuthorPhotos/Linhao.jpg}}]{Lin Hao}
% A
% \end{IEEEbiography}
% \begin{IEEEbiography}
% % [{\includegraphics[width=1in,height=1.25in,clip,keepaspectratio]{./AuthorPhotos/MustafaKishk.jpg}}]{Mustafa A. Kishk} 
% (Member, IEEE) received the
% B.Sc. and M.Sc. degrees from Cairo University
% in 2013 and 2015, respectively, and the Ph.D. degree
% from Virginia Tech in 2018. He is currently a
% Post-Doctoral Research Fellow with the CTL at
% KAUST. His current research interests include stochastic geometry, energy harvesting wireless networks, UAV-enabled communication systems, and
% satellite communications.
% \end{IEEEbiography}

\end{document}